\documentclass[reqno]{amsart}
\usepackage{graphicx}

\usepackage{enumitem}
\usepackage[numbers,sort&compress]{natbib}
\usepackage[foot]{amsaddr}
\usepackage[final,color,notref,notcite]{showkeys}
\usepackage{booktabs}
\usepackage{dsfont}
\usepackage{xcolor}
\usepackage{colortbl}
\usepackage{multirow,array}

\usepackage{hyperref}
\hypersetup{colorlinks,citecolor=blue}
\usepackage[inline]{asymptote}
\definecolor{labelkey}{rgb}{0,.56,.7}
\definecolor{tablerows}{rgb}{0.906, 0.8, .949}

\usepackage[charter]{mathdesign}

\theoremstyle{plain}

\theoremstyle{definition}

\newtheorem*{exa}{Example}

\theoremstyle{remark}

\def\bphi{\boldsymbol{\phi}}
\let\ox\otimes

\mathchardef\myhyph="2D

\DeclareMathAlphabet{\pazocal}{OMS}{zplm}{m}{n}   %  mathcal

\newcommand{\Hcal}{\pazocal{H}}

\def\bbZ{\mathbb{Z}}
\def\bbR{\mathbb{R}}

\DeclareSymbolFont{Eulerscripteusm10}{U}{eus}{m}{n}
\SetSymbolFont{Eulerscripteusm10}{bold}{U}{eus}{b}{n}
\DeclareMathSymbol{\euB}{\mathord}{Eulerscripteusm10}{"42}
\DeclareMathSymbol{\euC}{\mathord}{Eulerscripteusm10}{"43}
\DeclareMathSymbol{\euF}{\mathord}{Eulerscripteusm10}{"46}
\DeclareMathSymbol{\euI}{\mathord}{Eulerscripteusm10}{"4A}
\DeclareMathSymbol{\euK}{\mathord}{Eulerscripteusm10}{"4B}
\DeclareMathSymbol{\euR}{\mathord}{Eulerscripteusm10}{"52}
\DeclareMathSymbol{\euS}{\mathord}{Eulerscripteusm10}{"53}
\DeclareMathSymbol{\rH}{\mathord}{Eulerscripteusm10}{"48}

\newcommand{\nn}{\nonumber}

\def\df{\overset{\mathrm{df}}{=}}
\newcommand{\ket}[1]{\mathop{|#1\rangle}\nolimits}

\newcommand{\kbr}[2]{| #1\rangle\!\langle #2 |}
\newcommand{\diag}{\mathop{{\mathrm{diag}}}}

\newcommand{\ceil}[1]{\left\lceil#1\right\rceil}

\newcommand{\id}{\mathop{{\mathrm{id}}}\nolimits}
\newcommand{\di}{\mathop{{\mathrm{di}}}\nolimits}

\def\a{\alpha}

\def\g{\gamma}
\def\d{\delta}

\def\om{\omega}

\def\s{\sigma}

\def\ups{\upsilon}
\def\vt{\vartheta}

\def\om{\omega}

\allowdisplaybreaks

\begin{document}

\title{Efficient and scalable inter-module switching for distributed quantum computing architectures}

\author{Kamil Br\'adler}

%\address{}

%\date{\today}

\begin{abstract}
Large-scale fault-tolerant quantum computers of the future will likely be modular by necessity or by design. Modularity is inevitable if the substrate cannot support the desired error-correction code due to its planar geometry or manufacturing constraints resulting in a limited number of logical qubits per module. Even if the computer is compact enough there may be functional requirements to distribute the quantum computation substrate over distant regions of varying scales. In both cases, matter-based quantum information, such as spins, ions or neutral atoms, is the most conveniently transmitted or mediated by photonic interconnects. To avoid long algorithm execution times and reduce errors, each module of a universal quantum computer should be dynamically interconnected with as many other modules as possible. This task relies on an optical switching network providing any-to-any or sufficiently high simultaneous connectivity. In this work we construct several novel and decentralized switching schemes based on the properties of the Generalized Mach-Zehnder Interferometer (GMZI) that are more economic and less noisy compared to commonly considered alternatives while achieving the same functionality. We show that as the number of logical patches $n$ grows the active optical depth scales as $\Theta(\log{n})$ without sacrificing the logical qubit connectivity. 
\end{abstract}

\maketitle

%\begin{asydef}
%import graph;
%import settings;
%\end{asydef}

\thispagestyle{empty}

\section{Introduction}\label{sec:intro}

Recent years witnessed astonishing  hardware advances in the field of fault-tolerant quantum computing (FTQC)~\cite{reichardt2024logical,2025Natur638920G,rodriguez2024experimental,ryan2024high,valentini2024demonstration,bluvstein2024logical,reichardt2024demonstration}  following more than two decades of theoretical breakthroughs. The incessant push of full-stack quantum computing companies developing different physical platforms to outdo each other leads to the inevitable: an exponential improvement of the quality of the components and entire computational modules across the board. A computational module is a physical substrate consisting of a collection of physical (raw) qubits interacting with each  other in a controlled way. It can be either a `patch' supporting a logical qubit encoded in a quantum error correction code (note that our use of the word patch does not imply any geometric constraint such as a 2D substrate) or parts thereof. The computational modules are, however, quite often limited in size. The reason may be  immaturity of a physical platform (typically limited by the chip size for the integrated platforms) or its extensive classical supporting infrastructure such as control electronics or a cooling system.  It has been recognized  early on that quantum computers will be distributed~\cite{monroe2014large} (in a data center or even on a larger scale) and so the computational substrates will have to be linked to act as a single computational entity. Some platforms do promise an extreme qubit density with an overall small footprint~\cite{amitonov2024spin,burkard2023semiconductor,gonzalez2021scaling} but it is unlikely that the first generation of large-scale FTQCs is going to be built on a single substrate.

One of the few reasonable strategies to connect physically separated modules is by photonic interconnects. The minimal requirement  a quantum system  must satisfy in order to realize it is to be able to generate an entangled photon-matter pair, where the photon is at visible or near-infrared frequencies and `matter' is a physical system carrying quantum information (e.g. electron spin, atom, ion or a quantum dot~\cite{simmons2024scalable,de2024spin,rodriguez2024experimental,dasu2025breaking,reichardt2024logical,henriet2020quantum}). The photon is then either sent to a linear-optical module, where it interacts with another photon from the second module by means of a linear-optical transformation (e.g. one of many variants of the Bell state measurement~\cite{barak2007quantum,lim2006repeat} also called a fusion~\cite{browne2005resource})~\cite{simmons2024scalable,de2024spin}, or directly to the second module, where it is deterministically entangled with another matter qubit~\cite{bhaskar2020experimental}. In both cases,  the result is  two matter qubits entangled. Another type of physical platforms, where  photonic interconnection is a natural option, is linear-optical quantum computing~\cite{barrett2010fault,auger2018fault,li2010fault,bartolucci2023fusion,pankovich2024high}. Unlike the matter-qubit interaction, the photons themselves are the carriers of quantum information. One can again use linear or non-linear quantum optical gadgets to entangle (typically) large photonic states generated by physically separated substrates.

If we had just two hardware modules its interconnection wouldn't really be a big issue from the quantum architecture point of view. The difficult task is when we have to connect many modules with each other, each one typically  supporting one or more logical qubits in the form of a quantum error-correction (QEC) code. How exactly do we implement a multiple and simultaneous optical interconnection between any pair of modules? The literature studying distributed quantum architectures~\cite{caleffi2024distributed} for various platforms is often scant on details by simply referring to any-to-any or high-connectivity optical switching networks~\cite{monroe2014large,simmons2024scalable,weaver2025scalable,sutcliffe2025distributed,sunami2025scalable,awschalom2021development}. Given a high number of connections in the  large-scale fault-tolerant quantum computer such a device deserves a lot of attention. It must be versatile enough to be able to simultaneously route many photons from different modules to common entangling areas (for the Bell state measurement~\cite{brown2020universal} or other entangling linear-optical protocols) or directly connecting the substrates -- all this without considerably contributing  to the overall error budget and excessive operational\footnote{Heavy quantum traffic can be alleviated by, for example, temporal or frequency multiplexing but we will not explore these strategies here~\cite{bartolucci2021switch,bartolucci2023fusion}.} or manufacturing overhead. As a black-box device, the desired high connectivity can be achieved by a `non-blocking switch' whose properties will be recalled in the next section.

The requirement of the constrained error budget is absolutely crucial. It is not a problem to create a switching network of any size and functionality by concatenating many  switches. But each (typically) integrated switch has a certain amount of passive and active layers and in- and out-couplers that are among the biggest sources of  photon loss. So our primary goal is to  investigate how to achieve full or very high simultaneous connectivity while keeping their number constant and smaller than state-of-the-art. We present several  switching schemes, where the main role is played by the Generalized Mach-Zehnder Interferometer (GMZI)~\cite{lagali2000generalized}. We show that despite the GMZI's limited switching capabilities, earning it the name `blocking switch' (to be explained in the next section as well), one can achieve the same switching functionality in a more efficient and less noisy way than the best schemes based on Spanke's network~\cite{spanke1986architectures}, which boasts a constant active depth if also the GMZIs  serve as its building block~\cite{bartolucci2021switch}.  Given the recent extraordinary manufacturing advances of (really) ultra-low loss and large-scale GMZIs~\cite{psiquantum2025manufacturable} we believe that they can be deployed in the very near future as well as on a large scale.

Both entanglement approaches (probabilistic and direct) are  affected by errors in the form of photon loss and, whenever applicable, by poor photon distinguishability impeding perfect photon interference. How exactly these errors affect the performance depends on whether the photons are intended to carry quantum information  as already mentioned. If it is the case, photon loss is particularly damaging since it is often unheralded and it quickly degrades the properties of the final photonic computational substrate (the QEC code parameters such as its threshold). If photons are not information carriers none of these error mechanisms damages quantum information directly but they often slow down the rate of matter entanglement. This is because the entanglement success rate through photon interaction is  affected by errors and photon loss in particular. The success probability of, for example, the perfect Bell measurement (without any encoding or boosting) can't exceed 50\%\footnote{Note that microwave photons have the advantage of easy access to non-linear interactions resulting in the deterministic Bell measurement but even superconducting platforms seriously consider two-way transduction in order to link two remote substrates~\cite{ang2022architectures}.} and photon loss reduces it further.
\begin{figure}[t]
  \resizebox{11.5cm}{!}{\includegraphics{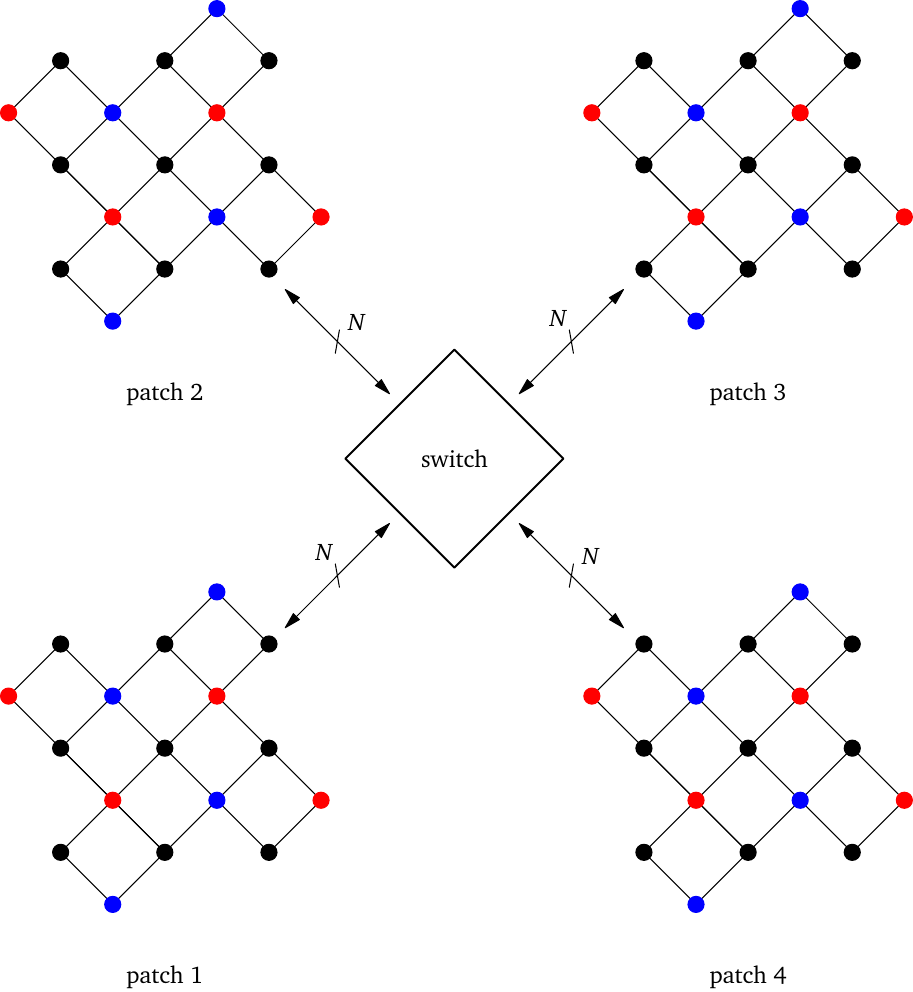}}
  \caption{A high-level picture of simultaneous any-to-any (sa2a) connectivity studied in this paper. Any pair of logical modules (four surface code patches for illustration) are required to be simultaneously connectable by $N$ links, typically optical fibers. This corresponds to three perfect matchings: $[(1,2),(3,4)], [(1,3),(2,4)]$ and $[(1,4),(2,3)]$. }
  \label{fig:sa2a}
\end{figure}
The main result in the form of several novel photonic-based simultaneous any-to-any (sa2a) switching schemes for modular quantum computing architectures is presented in Sec.~\ref{sec:GMZIswitching}. We slash all important figures of merit -- most notably the active depth and the number of fiber-to-chip couplers. It is preceded by Sec.~\ref{sec:switchingNetworks}, where we introduce some of the general properties of optical switches whose development and terminology predates even the era of classical optical communication. A thorough technical analysis supporting our main result is presented in Appendix~\ref{sec:GMZImath}, where we rigorously study and describe the GMZI as the principal optical component. A powerful formalism of the  Wigner $d$-matrices (not to be confused with the Wigner distribution function) allows us to exhaustively characterize the quantum switching properties of a practically important class of GMZIs. Appendix~\ref{sec:benchScale} concludes with a toy but nontrivial example of how the GMZI resources and optical depth scale with a increasing number of interconnected patches of the rotated surface code.

\section{Photonic switching networks}\label{sec:switchingNetworks}
\subsection*{The purpose of routing light}

We can distinguish between two main scenarios where the photons have to be routed with the help of optical switches. In the first case, there is a designated group of senders and receivers and the goal is to be able to send quantum optical states from any sender to any receiver, preferably simultaneously involving many senders and receivers. An optical switch with this property is called \emph{non-blocking}. In the second case, illustrated in Fig.~\ref{fig:sa2a} (with the $[9,1,3]$ rotated surface code~\cite{bravyi1998quantum,horsman2012surface} chosen for illustration), there is no special group of senders and receivers and the task is to be able to route single photons or other quantum-optical states from, say, any physical substrate to any other substrate. We will call this the \emph{simultaneous any-to-any} (sa2a) connectivity which owes its name to two crucial requirements:
\begin{enumerate}
  \item Any pair of modules or substrates must be connectable,
  \item Connecting a pair doesn't block any other pair from being connected at the same time.
\end{enumerate}
In this work we will be concerned mostly with the sa2a connectivity but at the heart of both switching tasks lies a single black box device with $N$ inputs and $M$ outputs: essentially a programmable optical permutation network. The difference is how this $N\to M$ device, which may be blocking like GMZI or non-blocking like Spanke's network\footnote{One should not be confused by the fact that Spanke itself can be implemented with GMZIs~\cite{bartolucci2021switch}.}, is used. Also note that the senders/receivers scenario (no matter what type of switch is used) may be thought of belonging more to quantum communication than computing but there is a plenty of use cases in the quantum computing world too. For example, there may be a central magic state factory (senders), which uses an optical switch to deliver the magic states to different QEC patches (receivers). CSS stabilizer measurement is another example. We will very briefly visit this topic at the end of Sec.~\ref{sec:GMZIswitching} for the sake of completeness.

Although the following discussion is valid  both for classical and quantum light our main goal is to describe how to route highly non-classical states such as single photons and other Fock states. That being said, in the context of this paper, the non-classical character of Fock states will only manifest itself by the appearance of a phase accompanying a basis state spanning the corresponding multiboson Fock space. We rigorously show that the GMZI, despite putting the incoming photons through complicated multiphoton interference, treats them \emph{almost} like classical (distinguishable) particles. More precisely, the GMZI in the switching regime is completely insensitive to the relative time of arrival of the photons: whether or not two or more photons arrive at the same time the routing operation succeeds. The photons are routed through  the GMZI in the same order they enter and  it is only the phase behavior the boson character of photons shows itself.   Even though a realistic component description goes beyond the scope of this paper this observation may have important ramifications for the practical use of the GMZIs, making them more forgiving (in fact, completely oblivious) to the issue of photons from different sources typically plaguing photon interference experiments.

\subsection*{Black box non-blocking switch}

\begin{figure}[t]
  \resizebox{8cm}{!}{\includegraphics{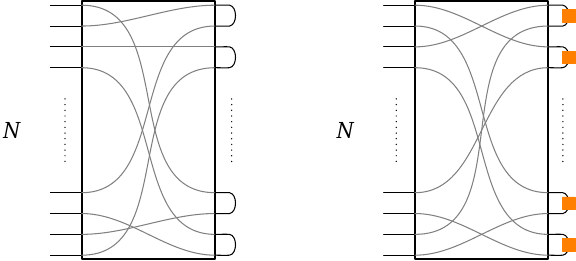}}
  \caption{The output ports of an $N\to N$ non-blocking switches (the rectangle black boxes) are hardwired so is possible to convert the sender/receiver operation mode to the sa2a scenario by simultaneously connecting $n=N/2$ pairs of the output ports either deterministically  (on the left) or  probabilistically (on the right), where the orange box is a general entangling operation such as the BSM. The grey lines illustrate a particular switching configuration.}
  \label{fig:all2all}
\end{figure}

As a black box, an $N\to M$ non-blocking switching network implements all possible permutations, thus  simultaneously directing up to $\min{[N,M]}$ input photons from any subset of input ports to any subset of output ports. The number of switching configurations is
\begin{equation}\label{eq:numberPerms}
  p=\prod_{i=0}^{\min{[N-1,M-1]}}(M-i)=\min{[(M)_M,(M)_N]},
\end{equation}
where $(x)_n\df x(x-1)\dots(x-n+1)$ is the falling factorial. Crucially, a device with this functionality can be turned into a device providing sa2a connectivity both for the direct and probabilistic entanglement generation by hardwiring pairs of the output ports, cf.~Fig.~\ref{fig:all2all}. We hardwire neighboring output ports but any pair-hardwiring choice is valid since the device can realize any permutation. In this operational mode there is no distinction between the  senders and receivers. Instead, each participant is associated with one of the $N$ input ports and sa2a demands $M=N$. Assuming $N=2n$, the number of simultaneous connections (pairings) is~$n$ and the number of such configurations is the  number of perfect matchings $\#\mathrm{pm}(N)=N!!=(2n)!!=(2n)!/2^n/n!$, where $!!$ denotes the double factorial. Note that quantum networks with $N$ participants often require much less connectivity than $\#\mathrm{pm}(N)$ but we believe that for distributed quantum computing the bigger connectivity will always be better (especially when the logical modules support a limited number of logical qubits and so many modules are required for large-scale quantum algortihms) .

The setup from Fig.~\ref{fig:all2all} achieves the desired goal even in the case where each substrate needs to be connected to any other by not one fiber but, say, by $k$~fibers. This will typically be the case. But then we don't even need a switching network implementing the full symmetric group $|S_{kN}|=(kN)!$ since $k$~fibers from one substrate have to  be always  collectively routed to another substrate. There is a potential for significant savings of resources by not having to implement all permutations and our proposal falls into this category.

\subsection*{Standard optical non-blocking switch constructions}

The number of non-blocking optical switch proposals is vast and they vary in their properties and complexity. Increasingly sophisticated networks implementing the non-blocking black box switch were invented~\cite{clos1953study,benevs1962algebraic} and later adopted for classical and quantum optical communication~\cite{spanke1987n,lagali2000generalized,padmanabhan1987dilated}. They differ in the number and complexity of active components such as a programmable beam-splitter or the number of crossings one needs to be aware of in 2D integrated photonic platforms. It is not our goal to list them all~\cite{lee2018silicon,cheng2018recent} (some explicit examples can be seen e.g. here~\cite{li2024high,li2016hierarchical}) and so  we will quickly zoom in on the scheme most relevant to the topic of this paper. For us, the \emph{active depth}, that is, the number of active optical components a single photon has to pass through and the number of chip-to-fiber couplers per photon are the main figures of merit we would like to minimize. These are the biggest contributors to photon loss in switching networks in the case of integrated photonics. There are free-space alternatives for optical switching such as the MEMS (the Micro-Electro Mechanical Systems -- moving mirrors), which offer a large number of input and output ports together with a low loss regime, but the mechanical nature of the switching substrate makes them very slow and therefore unsuitable for large-scale FTQC.

%The active depth can be decreased to a linear scaling but, in general, the fiber crossings are always present.

\begin{figure}[h]
  \resizebox{13.5cm}{!}{\includegraphics{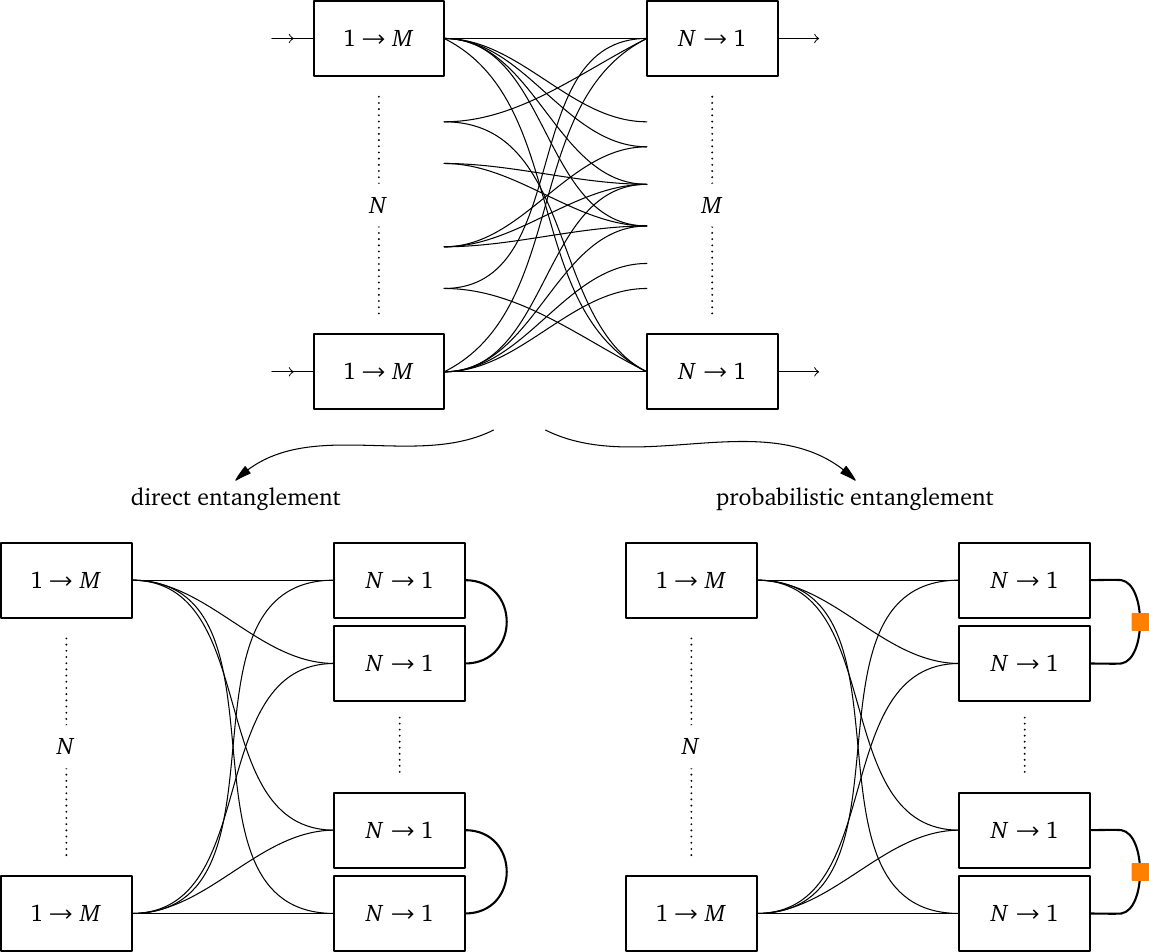}}
  \caption{(Top) $N\to M$ Spanke's (also called switch-and-select) network as an example of the switching device capable of the routing illustrated in Fig.~\ref{fig:all2all} for $M=N=2n$ with the hardwired output ports. The hardwiring depends on the available entanglement strategy (bottom left for direct and bottom right for probabilistic). If the  $1\to M$ and $N\to1$ switches were realized by concatenated binary trees Spanke's network would be lossy and currently unsuitable for large-scale FTQC. Here we consider them to be realized with GMZIs~\cite{bartolucci2021switch} and hardwired -- thus achieving a constant active depth four. The purpose of this paper is to benchmark the GMZI-based Spanke switches against our GMZI-based proposals.}
  \label{fig:spanke}
\end{figure}

The most straightforward (and the least suitable) option to make a non-blocking switch  is to program one of many universal optical unitary $U(N)$ networks~\cite{reck1994experimental} often built by concatenating Mach-Zehnder interferometers (MZIs) such that it implements the desired permutation matrix. This option is not scalable since the active optical depth grows quadratically with the number of input modes and it is currently infeasible to manufacture ultra-low loss integrated circuits with active elements growing that fast at the level necessary for FTQC. One can use specialized programmable circuits instead which only implement  a subset of the unitary group  $U(N)$ necessary for switching, which is the symmetric subgroup $S_N$. This dramatically decreases the active depth to $\pazocal{O}(\log{N})$ in some cases. One of them is $N \to M$ Spanke's switch~\cite{spanke1986architectures}  (it sometimes goes under the name `switch and select'~\cite{cheng2018recent}) explicitly used or manufactured e.g. here~\cite{main2025distributed,chen2012compact}. Its most common  realization  requires  $1\to M$ and $N\to1$ switches originally~\cite{spanke1986architectures} obtained by concatenating $\log{M}$ and $\log{N}$ elementary $1\to2$ or $2\to1$ switches (see Fig.~\ref{fig:spanke} where the boxes would be such binary trees). In modern integrated optics these elementary switches are MZIs or  strings of multi-ring resonators~\cite{nikolova2017modular} (MRRs), which is essentially an asymmetric binary tree.

However, the logarithmic active depth is still not good enough at the single-photon level used in photonic-based or assisted FTQC. To get to a constant active depth for any $N,M$, one can use the GMZIs mentioned earlier instead of concatenated MZIs~\cite{bartolucci2021switch}. Similarly to the standard MZI, that  can be programmed such that it acts as a $2\to2$ switch with just one active layer, the GMZI can be programmed to provide a limited routing device (see the next section about how exactly limited it is) with any $N$ input and $M$ output ports. A single photon can be directed from any input to any output port with, again, just one active layer per photon. So if we substitute  all MZI binary trees in the $N\to M$ Spanke network by a suitable GMZI we reduce the active optical depth to two~\cite{bartolucci2021switch} as depicted in Fig.~\ref{fig:spanke} and one can implement the schemes in Fig.~\ref{fig:all2all} which implies an active depth four (four per photon on the left or two per photon (but twice because two photons meet at the orange entangler) on the right).

We will use the sa2a switching schemes in Fig.~\ref{fig:all2all} implemented by GMZI-based Spanke networks depicted in Fig.~\ref{fig:spanke} as a protocol to be compared wit our proposal. Not only because it is also based on GMZIs but mainly it seems to be the state of the art for the active depth and chip-to-fiber number scaling.  In the next section we present our main results, where by deconstructing Spanke's network equipped with the GMZIs we achieve sa2a with  a smaller (constant) active depth and a smaller number of optical couplers while (as a bonus) also making the whole switching network truly distributed. Note that  despite the potentially distributed character of the non-blocking switch in Fig.~\ref{fig:spanke}  (that is, with the $1\to M$ and $N\to1$ GMZIs connected by fibers and not on the same substrate) it often figures as a central node~\cite{sutcliffe2025distributed} similarly to the scheme depicted in Fig.~\ref{fig:sa2a}. Depending on the practical circumstances of how a quantum computer is deployed it may or may not be advantageous to have a central switchyard but our novel schemes are fully distributed by design.

\section{GMZIs for efficient simultaneous any-to-any connectivity}\label{sec:GMZIswitching}

The main optical component we will use is the GMZI whose properties and functionality we briefly describe here but its detailed quantum-mechanical analysis of how it can be used for routing any quantum mechanical state of bosons can be found in Appendix~\ref{sec:GMZImath}. All switching schemes introduced in this section rely heavily on the results presented there.

The GMZI with $N$ input and $M$ output ports~\cite{lagali2000generalized} was designed to route a classical optical mode from any input to any output port with a constant active layer number equal to one. At the classical level, the routing is described by a transfer matrix~$T$. It is a product of three maps: an active layer of phase modulators sandwiched in between two passive optical networks called multi-mode interferometers (MMIs). The first network has $N$ input ports and the second network has $M$ output ports. By choosing a given input  (output) port it is possible to effectively create an $N\to1$ or $1\to M$ switching network as originally intended. In this way, other ports would have been blocked (unusable) until the routing was over and for this reason it is sometimes called an  $N\to M$~\emph{blocking switch}. By studying the symmetry properties of the transfer matrix~$T$ it was, however, further deduced~\cite{lagali2000generalized} that $T$ acts as a certain permutation of all input ports which is determined by the configuration of the phase shifters. This led to the conclusion that an $N\to M$ GMZI can route up to $N (M)$ optical modes simultaneously for $N\geq M (N\leq M)$. Indeed, if for a fixed collection of phases the matrix $T$ acts as $T:k\to j_k$ for  $1\leq k\leq\min{[N,M]}$ such that $T_{k,j_k}=e^{i\phi}$ and zero otherwise, then limited simultaneous classical transmission is possible. This does not mean that the GMZI becomes a fully non-blocking switch. The number of suitable phase configurations is far smaller than it is necessary to realize the action of whole symmetric group.

How does the GMZI act as a quantum device, that is, if  we use it as a switch for single- or multi-photon quantum states? One can describe the GMZI quantum-mechanically as a unitary $U$, being a product of three other unitaries: $U=VDW$, where $V,W$ are  linear-optical circuits and  $D$ is a diagonal phase matrix. The Hilbert  space they act on must be a proper multiphoton Fock space necessary to describe the evolution of a multiboson quantum state. Note that at the quantum-mechanical level, the operators $V,W$ are responsible for quite complicated multi-photon interference effects generalizing the HOM effect two photons experience at a beam splitter~\cite{hong1987measurement}. What is then the relation between the classical transfer matrix~$T$ and the unitary matrix $U$? The authors of~\cite{bartolucci2021switch} argue that they are the same. That raises a lot of questions since the classical and quantum operators act on wildly different spaces. For example, the diagonal phase operator $D$ is certainly not acting on just $M$ or $N$ classical modes but rather on (often) a high-dimensional $N(M)$-mode Hilbert space as we will see. On the other hand, the transfer matrix is a unitary matrix (it is a permutation after all) so could it be that in the end the complicated action of $U$ on an arbitrary input Fock state reduces to that?

To clarify this somewhat opaque situation we describe the GMZI fully quantum-mechanically without referring to its classical origin. We  focus on the case of the $N\to N$ GMZIs where $N=2^k$, $V=W^\dagger$ and $W$ is the well-known quantum Fourier transform~\cite{barak2007quantum}. We denote $\euS(\bphi)\df W^\dagger D(\bphi)W$ the unitary switching operator for the proper choice of an $N$-tuple~$\bphi$. This is a special case of a general $N\to M$ GMZI switch  but it is also a case of a great practical importance\footnote{Whenever we write $N\to M$, where $N$ or $M$ is not  a power of two, we tacitly assume such an GMZI to be the one with $N,M$ `rounded up' to the closest power of two.}. We fully characterize the action of such devices and describe how to find the switching transformation $\euS_{\bphi}$ as a function of $\bphi$ for any $N$, where $\phi_i=\{0,\pi\}$. Note that not all combinations of $0$'s and $\pi$'s in $\bphi$ are possible and we also determine which one are the legitimate switching phase configurations.  To give up the main technical result from  Appendix~\ref{sec:GMZImath} right away, we confirmed the close relationship between $T$ and $U$ but with an important twist related to the quantum character of the routed boson states. The switches' action reads~\eqref{eq:switchActGeneral}
\begin{align}\label{eq:switchActGeneralIntro}
  \euS_{\bphi}(\ket{0,\dots,n_i,\dots,n_j,\dots,0})
   &= (-)^{n_{\text{tot}}\bphi(N)/\pi}\ket{0,\dots,n_{\s^{-1}(i)},\dots,n_{\s^{-1}(j)},\dots,0},
\end{align}
where $n_{\text{tot}}$ is the total input/output photon number, $\s$ is a permutation as a function of~$\bphi$ and $\bphi(N)$ is the $N$-th (last) phase angle.

Our second main technical result from  Appendix~\ref{sec:GMZImath} is a derivation of the mapping (a bijection) between $\bphi$ and $\s$ accompanied by the characterization of which $\bphi$'s are valid switching configurations (see an example in~\eqref{eq:ex1cont}). We show in Table~\ref{table:N4example} all switching configurations of the $4\to4$ GMZI.  Clearly, the elements of the same sign form the Klein four-group $K_4\simeq\bbZ_2\times\bbZ_2$.
\begin{center}
    \renewcommand{\arraystretch}{1.2}
    \extrarowheight=\aboverulesep
    %\addtolength{\extrarowheight}{\belowrulesep}
    \aboverulesep=0pt
    \belowrulesep=0pt
    \begin{table}[h]
           \begin{tabular}{@{}>{\columncolor{white}[0pt][\tabcolsep]}  *{3}c @{}}
           \toprule
            {\cellcolor{lightgray}\ $\bphi$ } & $\s$ &   $(-)^{n_{\text{tot}}\bphi(N)/\pi}$    \\
                        \midrule
             {\cellcolor{lightgray}\ $(0000)$}  &   $()$  &  $+$  \\
             {\cellcolor{lightgray}\ $(\pi\pi\pi\pi)$}      &   $()$  &  $-$  \\
             {\cellcolor{lightgray}\ $(\pi0\pi0)$}      &   $(12)(34)$  &  $+$   \\
             {\cellcolor{lightgray}\ $(0\pi0\pi)$}  &   $(12)(34)$  &  $-$  \\
             {\cellcolor{lightgray}\ $(\pi\pi00)$}  &   $(13)(24)$  &  $+$  \\
             {\cellcolor{lightgray}\ $(00\pi\pi)$}  &   $(13)(24)$  &  $-$  \\
             {\cellcolor{lightgray}\ $(0\pi\pi0)$}  &   $(13)(24)(12)(34)=(14)(23)$  &  $+$  \\
             {\cellcolor{lightgray}\ $(\pi00\pi)$}  &   $(14)(23)$  &  $-$  \\
             \bottomrule
             \hline
            \end{tabular}\\ \vskip .3cm
    \caption{All allowed quadruples of phases $\bphi$ for the $4\to4$ GMZI including the resulting permutations $\s$ and phases from Eq.~\eqref{eq:switchActGeneralIntro} if $n_{\text{tot}}$ is odd. The phases are trivial for any even $n_{\text{tot}}$ and $\bphi(N)$ is the last digit of $\bphi$.}
    \label{table:N4example}
    \end{table}
\end{center}

In the following text we present a number of new switching schemes based on GMZIs and fully exploit the characterization result. Our goal is to show its advantageous properties compared to the GMZI-based Spanke's network while achieving the same functionality. Some of our general constructions will be accompanied by small examples with a more detailed description based on the technical results presented in Appendix~\ref{sec:GMZImath}. It is followed by two case studies of quantum computing architectures for photon-mediated spin entanglement of matter qubits~\cite{simmons2024scalable} (also applicable to other proposals such as~\cite{de2024spin,nickerson2014freely,nemoto2014photonic}) and modern linear-optical quantum computing~\cite{pankovich2024high}. In particular, we show how to implement a transversal CNOT and the merge lattice surgery operation~\cite{horsman2012surface,vuillot2019code} in a probabilistic entanglement setting~\cite{barrett2005efficient,lim2006repeat}.

Note, however, that the common low-level denominator of many photon-mediated protocols is a distribution and efficient routing of entangled states to implement some form of teleportation enacting a two-qubit logical gate between two remote matter qubits in the sa2a spirit. The GMZI-based switching scheme developed in this work is thus  not limited to  the two protocols we are going to present here but rather applies to a large class of probabilistic entanglement strategies via photonic interconnects~\cite{li2016hierarchical,awschalom2021development,ang2022architectures,sunami2025scalable,li2024high,sinclair2025fault,stephenson2020high,
wan2019quantum,hahn2025deterministic,ataides2025constant,shalby2025optimized,avis2023analysis,pattison2024fast,nigmatullin2016minimally,weaver2025scalable,kalb2017entanglement,
marton2025lattice,ruskuc2025multiplexed,sakuma2024optical,jacinto2025network}.

\subsection*{Novel GMZI-based sa2a switching schemes and their general properties}

Let's put our GMZI characterization results to use. The switching proposals based on this analysis are applicable to any matter-based quantum computing platform assisted by optically mediated entanglement (by a photon emission process or by entangling a photon from an external source and `reflecting' it~\cite{bhaskar2020experimental})  and subsequently collecting the emitted/reflected photon state to (typically) an optical fiber. We can further divide these architectures into two groups: (i) the matter qubits of a single logical module can  interact with each other through direct coupling (such as the neutral atoms and trapped ion-based platforms~\cite{reichardt2024logical,bluvstein2024logical,reichardt2024demonstration}) and so the emitted photons serve solely for the inter-module interaction and (ii) the emitted photons also mediate the matter-qubit interaction within the module itself~\cite{simmons2024scalable,de2024spin,nickerson2014freely}.

Note that if the direct coupling within a module is not possible, the ideas for GMZI-based switching presented here are directly applicable to the intra-module connectivity as well, especially if the module supports more than one logical qubit like in modern code families~\cite{malcolm2025computing,bravyi2024high}. In that case, the intra- and inter-module connectivity cannot be dealt with separately. This is an important topic that goes, however, beyond the scope of this paper. For now we only note that if, for example, each module supports tens of high-distance logical qubits and there are many such modules then every switching network enabling sa2a would quickly reach its practical limits by trying to link an arbitrary logical qubit from any module with any other module's logical qubit.

\subsection*{GMZI sa2a switching for direct entangling protocols}

\begin{figure}[t]
  \resizebox{9cm}{!}{\includegraphics{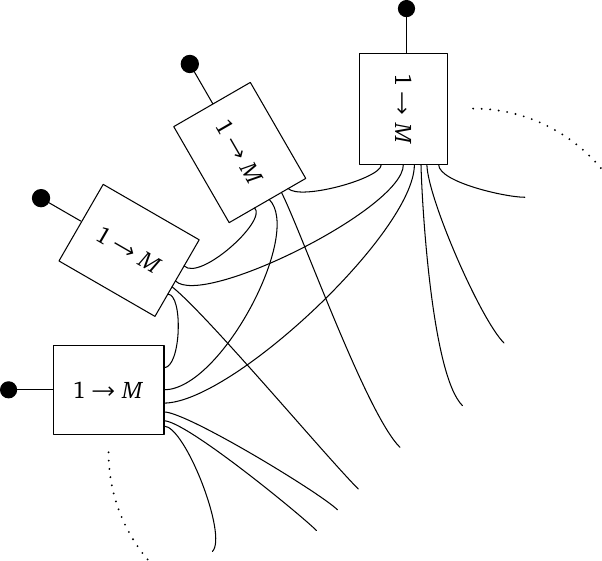}}
  \caption{A collection of $N=M+1$ matter qubits (the black dots), each emitting a single photon to a fixed input port of a $1\to M$ GMZI. By counting the components every photon state has to go through, our scheme is simpler and less noisy than Spanke's network in Fig.~\ref{fig:spanke} (bottom left), which is also built from GMZIs and with an equivalent switching functionality. }
  \label{fig:GMZIdirec1toN}
\end{figure}

Our first and the simplest scheme on the way to implement an inter-module logical Clifford operation, where nevertheless the modules are not supporting logical qubits yet, is already going to outperform the reference GMZI-based Spanke's switch. To that end, assume that each module is just a single matter qubit capable of emitting and absorbing an $m$-photon Fock state  (typically $m=1$ in the photon-mediated quantum architectures) in one spatial mode and there are $N=2n$ modules ($N$ even is just for convenience) following the black box model in Fig.~\ref{fig:all2all} on the left. We wish to implement sa2a  just like it is possible with Spanke's network and so any pair of matter qubits must be possible to simultaneously entangle. The setup is depicted in Fig.~\ref{fig:GMZIdirec1toN}. We choose any pair of matter qubits and one from each pair will emit a photon. The $1\to M$ GMZIs, where $M=N-1$, direct all photons simultaneously to their destinations  and all pairs will be able to get entangled. Our scheme is manifestly better  in terms of how many GMZIs are needed compared to the GMZI-based Spanke switch on the bottom left of Fig.~\ref{fig:spanke} ($N-1$ vs $2N$) while at the same time halving  the active depth every emitted photon state has to pass through from four to two and the number of fiber-to-chip couplers from eight to four.

\begin{figure}[t]
  \resizebox{10cm}{!}{\includegraphics{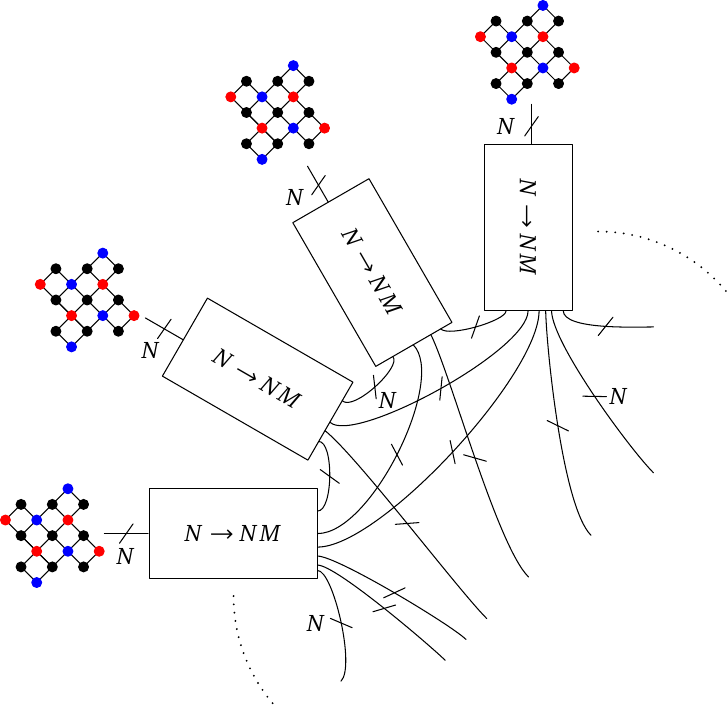}}
  \caption{A high-level depiction of a typical problem in modular quantum computing. A collection of $M+1$ logical qubits (here illustrated as rotated surface code patches) whose $N$~supporting physical qubits are required to interact to enact simultaneous two-qubit operations such a transversal logical CNOT. The physical qubits are sent  to their corresponding $N\to NM$ GMZIs and are routed simultaneously ($N$ and from all modules) to their destination logical module. See Fig.~\ref{fig:412direct} for a more detailed comparison the GMZI-based Spanke's switch.}
  \label{fig:GMZIswitch1multi}
\end{figure}

The switching advantage of our GMZI-based approach becomes yet more apparent when instead of a single matter qubit we have a true logical module supporting one or more logical qubits. To implement, for example, a transversal CNOT we may wish to simultaneously route several raw (data) qubits  to the same location and do the same for any other pair of logical modules simultaneously. Let~$N$ be a number of photons each module emits (it can be an arbitrary quantum-optical state, see Appendix~\ref{sec:GMZImath}). The situation is depicted in Fig.~\ref{fig:GMZIswitch1multi} and we see that we are using exactly the same circuit geometry as in Fig.~\ref{fig:GMZIdirec1toN}. The difference is a bigger size of all GMZI attached to the logical modules. Again we squarely beat the GMZI Spanke's switch. If we intended the same protocol with Spanke's network in Fig.~\ref{fig:GMZIdirec1toN} one option would be  to prepare $N$~copies of it and that is significant overhead.

\begin{figure}[t]
  \resizebox{14.5cm}{!}{\includegraphics{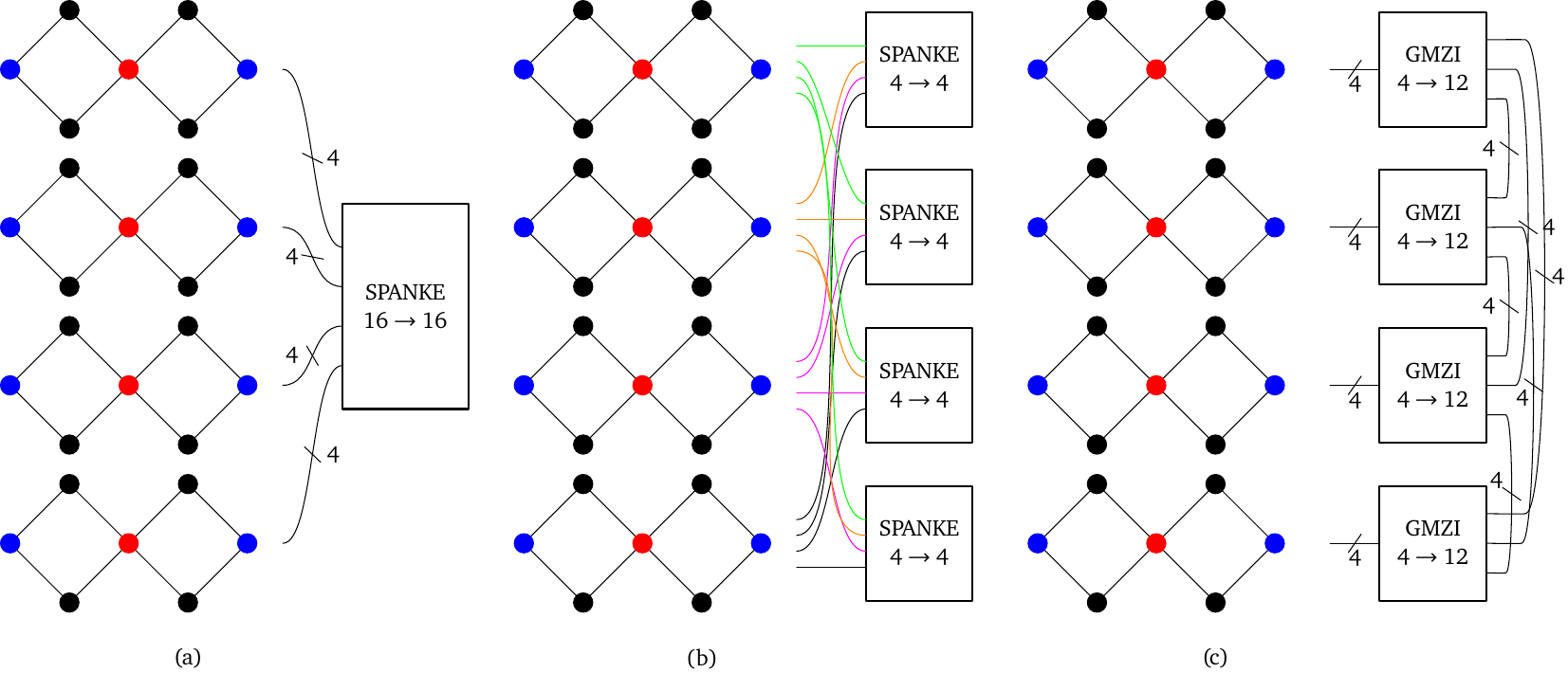}}
  \caption{A comparison of the GMZI-based Spanke network (a) and (b) reconnected as depicted in Fig.~\ref{fig:spanke} (bottom left) and our GMZI-based proposal (c) with the same functionality to enact a transversal CNOT between any possible two pairs of the $[4,1,2]$ surface code patches (simultaneously) for platforms allowing  direct entanglement of matter qubits and photons. The colors in (b) are for a more visible fibre tracking.}
  \label{fig:412direct}
\end{figure}

Let's elaborate on the exact comparison and show how to implement a transversal CNOT for $N=4$ patches of a smaller instance of the  surface code, namely the $[4,1,2]$ patch.  Each patch encodes one logical qubit and its stabilizer generators are $XXXX,IIZZ,ZZII$. The setup is depicted in Fig.~\ref{fig:412direct}.
Each code patch uses four data qubits and therefore to implement sa2a we must be able to simultaneously route four photons to $N-1=3$ different locations. This routing can be accomplished by a $16\to16$ GMZI-based Spanke's network hardwired like in Fig.~\ref{fig:spanke} (bottom left), which is made from 32 $1\to16$ GMZIs, the active depth of the circuit is four and the number of fiber-to-chip couplers is eight (see Fig.~\ref{fig:412direct}~(a)). If we compare it with our GMZI-based approach based on the general analysis developed in Appendix~\ref{sec:GMZImath} we can instead use only four $4\to12$ GMZIs, the active depth is two and the number of fiber couplers is four (see Fig.~\ref{fig:412direct}~(c)). This is quite a significant saving of resources and will have a profound impact on the loss budget. The reason for this favorable behavior is the property already apparent in the example of the $4\to4$ GMZI in Table~\ref{table:N4example}: the GMZI is capable of routing multiple input ports at once in different directions. So even though the switching subgroup of the symmetric group  is tiny (like the Klein group example) it does achieve interesting switching capabilities.

The comparison results in a heavy defeat for Spanke and to be fair one can come up with a better option illustrated in Fig.~\ref{fig:412direct}~(b) as well. We use four $4\to4$ GMZI-based Spanke's networks instead of one which brings the total resource count to 32 $1\to4$ GMZIs. This might be more competitive compared to four $4\to12$ GMZIs but the active depth as well the number of couplers remains unfavorably high  (four and eight, respectively).

Is there any advantage left to actually use the GMZI-based Spanke network? The author is not aware of any theoretical reason but there are practical limits for sa2a shared by both Spanke's and our approach. Going back to denoting GMZI as $N\to M$, the GMZIs as the building blocks for both schemes can't be reasonably well manufactured for an arbitrarily large number of ports. The issue is not a growing number $N$ of input ports (connections per a logical module) because one can just make $1\leq k\leq N$ GMZIs of size $N/k\to M$ if $N$ gets to high. The issue is the number of GMZI output ports $M$, which is essentially $M=mN$, where $m$ is the number of modules, and so $M$~becomes unrealistic. There are three solutions to this important practical problem: (i) Build more smaller $N/k\to mN/k$ GMZIs, (ii) one has to accept limited connectivity, that is, sacrificing the true (s)a2a which becomes essentially a quantum algorithm compilation problem, and, (iii) a switch concatenation strategy. For the last option the active depth increases but the GMZI strategy presented here will again gain an upper hand compared to the concatenated GMZI-based Spanke networks. We believe that the combination of all three strategies  can help build a large-scale distributed quantum computing fabric with a reasonable error budget.
\label{para:scaling}

\subsection*{GMZI sa2a switching for probabilistic entangling protocols}

Let's turn our attention to the probabilistic scheme in Fig.~\ref{fig:all2all} depicted on the right, where the photons from two computational modules are  entangled in a gadget, which is often  one of many variants of the Bell state measurement -- BSM (such as the un/partial/rotated Type-II fusion~\cite{browne2005resource}).  In that case, the success probability doesn't cross 50$\%$ (unassisted by ancillary photon states and considering no photo loss)  but the failure is heralded. The resource count of the GMZI-based Spanke's network is the same as in the previous, direct entanglement, see Fig.~\ref{fig:spanke} on the bottom right. To use our more efficient GMZI-based proposal we could modify the scheme in Fig.~\ref{fig:GMZIdirec1toN} or Fig.~\ref{fig:GMZIswitch1multi} if every fiber connecting any two GMZIs become equipped with its own entangling gadget. But that is an inefficient way of solving the problem because it leads to a quadratic growth of the number of entangling modules with the number of logical modules. To avoid this issue we introduce two different, more economic, approaches.

\begin{figure}[t]
  \resizebox{10cm}{!}{\includegraphics{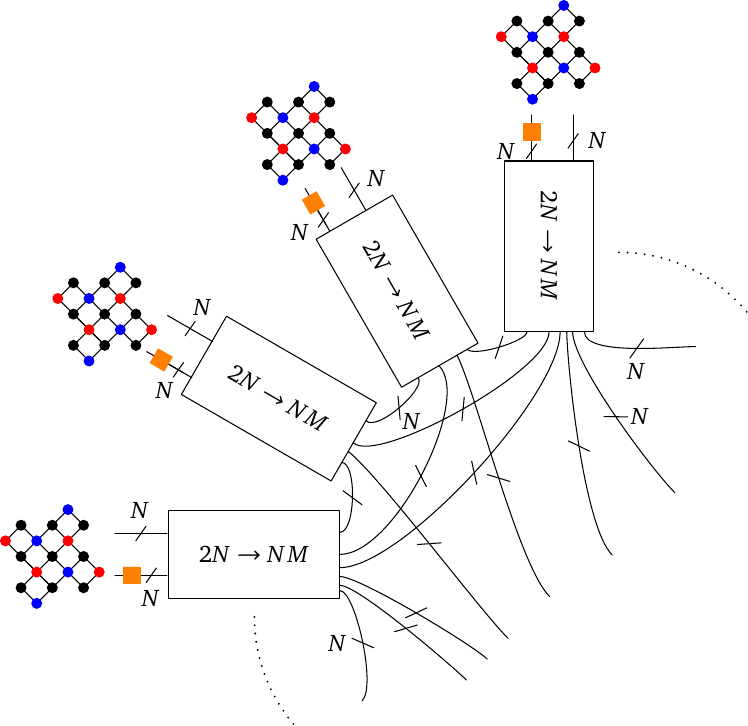}}
  \caption{By carefully inserting an entangling gadget (the orange box) between a logical module and an GMZI in probabilistic schemes we make sure that  in this way modules updated in this way are identical (equalized) and so we are not artificially splitting the communication network into a subset of senders and receivers. In this way, sa2a connectivity can be achieved by the GMZI-based switches similarly to the direct entanglement schemes. The only price to pay is an increased number of input ports from $N$ to $2N$ compared to the direct scheme in Fig.~\ref{fig:GMZIswitch1multi}.}
  \label{fig:GMZIswitch4}
\end{figure}

For the first proposal we notice that the entangling gadget can be inserted between a module and its GMZI. In this way, the number of entangling gadgets scales linearly with the number of logical modules. But we clearly cannot do this for all modules since every photon would, inconveniently, have to pass through two entanglers. We could divide the modules into two groups -- the senders equipped with the entangling modules and the receivers without them. But that would affect the desired sa2a connectivity since two receivers could not be linked. The solution is to equip all modules with  entanglers and the equal number of fibers connected to the same physical qubits as the entanglers. The scheme is depicted in Fig.~\ref{fig:GMZIswitch4} and what we are basically doing is some sort of `equalization' so that all modules look the same. Note that this process does not modify any  critical switching parameter such as the number of GMZI and the active depth which remains equal to two. It doubles the number the GMZI input ports compared to~\ref{fig:GMZIswitch1multi} from $N$ to $2N$ and also the actual physical qubits of each QEC code patch must be connectable to two external fibers instead of one.

We will these effects in the first case study following this section where we show how to implement a transversal CNOT in detail accompanied by an explicit GMZI switching configuration which is not obvious from the above figure. We showcase it together with a probabilistic entanglement setup suitable for the merge operation of lattice surgery~\cite{horsman2012surface} implementing a logical CNOT.

\begin{figure}[t]
  \resizebox{11cm}{!}{\includegraphics{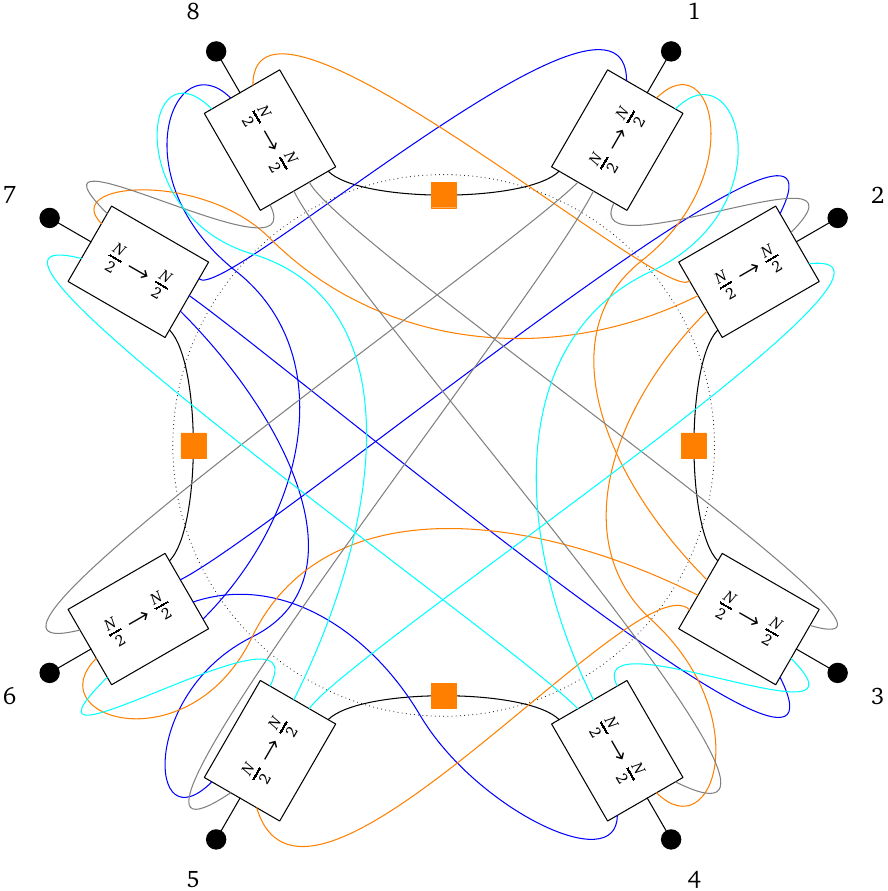}}
  \caption{Connectivity for $N=8$ matter qubits (the black dots) can be achieved by only $4\to4$ GMZIs if we erase the difference between the input and output sides of the GMZI and allow some photons to enter through the input, where the matter qubits are typically generated. Sa2a connectivity is achieved by routing through two or three GMZIs (two active layers for the modules connected by an orange entangler and three active layers otherwise). Despite a slightly increased active depth we still beat Spanke's scheme with the same functionality and in addition we manage to use smaller GMZIs. Pairs of GMZIs share the same color of the connecting fibers but this is just to make the figure less messy.  This illustrative example is formalized by recasting it in the language of graph theory with the help of Fig.~\ref{fig:mixedGraph}.  }
  \label{fig:GMZIhalf}
\end{figure}

The second proposal goes in a different direction. We hardwire the entanglement gadgets to several of the neighboring GMZIs and use some creative switching as illustrated on the case of $N=8$ in Fig.~\ref{fig:GMZIhalf} implementing the black-box scheme in Fig.~\ref{fig:all2all} (right). This scheme is noteworthy since we need only eight $4\to4$ GMZIs instead of sixteen $1\to8$ for Spanke's switch in Fig.~\ref{fig:spanke} (bottom right). Equally importantly, note that the photons from substrates with the hardwired entangling gates have to only pass through one GMZI (so active depth two per entangling attempt as before) and the rest through 2+1=3 GMZIs (active depth three). This is on average still less than the necessary active depth four in Spanke's switching protocol.

\begin{figure}[t]
  \resizebox{12cm}{!}{\includegraphics{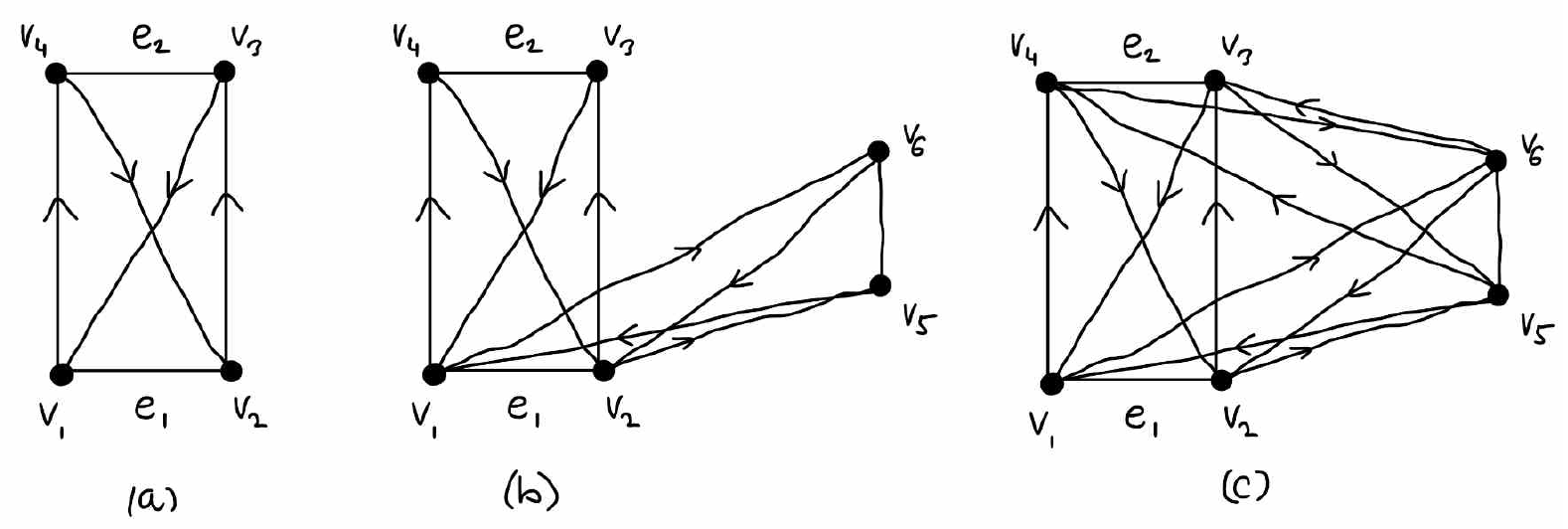}}
  \caption{A mixed graph $G$ constructed in the text.}
  \label{fig:mixedGraph}
\end{figure}

Fig.~\ref{fig:GMZIhalf} looks random and lacking structure so let's now formalize the result and show the scheme's validity for any even $N$ (not just $N=2^k$). Let $G$ be a trivial graph in the form of a collection of $N=2k$ vertices. We choose four vertices $v_i,i=1,2,3,4$ and attach to them two undirected edges $e_1,e_2$ such that the edges are disjoint (one of the three perfect matchings on four vertices, say, $e_1=(v_1,v_2),e_2=(v_3,v_4)$). We then attach four \emph{directed} edges connecting all vertices of $e_1$ with all vertices of $e_2$ such that the directed subgraph is oriented, see Fig.~\ref{fig:mixedGraph}~(a) for the resulting graph~$G$. We just constructed the $N=4$ version of the network in Fig.~\ref{fig:GMZIhalf} if we identify the vertices with four $2\to2$ GMZIs, each of them with a matter qubit substrate, where $e_1,e_2$ are the entangling modules and the directed edges indicate the direction of photons such that any pair of substrates can get entangled with at most three GMZI trips for both photons. Note that the differences of the in- and out-degrees  of all vertices of~$G$, $\d^+(v_i)$ and $\d^-(v_i)$ respectively, are zero.

Let's continue by showing that it holds for any even $N$. To this end, we add two more vertices $v_5,v_6$ to~$G$ and form a new undirected edge~$e_3=(v_5,v_6)$, see Fig.~\ref{fig:mixedGraph}~(b) for what happens next. We first connect $v_1,v_2$ with $v_5,v_6$ by directed edges such that each vertex has one ingoing and one outgoing edge. In this way we keep the in- and out-degrees of all vertices equal. Finally, we connect $v_3,v_4$ with $v_5,v_6$ in the same way as in  Fig.~\ref{fig:mixedGraph}~(c). As a result,  $\d^+(v_i)-\d^-(v_i)=0,\forall i$ and for each vertex there is an oriented cycle of length three, where one edge of each cycle is bidirectional. We can iteratively add new connected vertex pairs and the same procedure of adding directed edges to all previous vertices preserves $\d^+(v_i)-\d^-(v_i)=0$  creates new oriented 3-cycles. Indeed, in the next step after adding $v_7,v_8$  we would precisely obtain  the scheme in Fig.~\ref{fig:GMZIhalf} for $N=8$.

So we found a strictly better, fully decentralized switching device, which for any $N=2^k$ (the special case of interest) needs smaller GMZIs resulting in a reduced number of active and passive layers. As a consequence, any two pairs of substrates can be entangled in a simultaneous way going through at most 3 GMZIs while at the same time also decreasing the size of the used GMZIs from $N\to N$ to $N/2\to N/2$. As is the case of Fig.~\ref{fig:GMZIswitch1multi}, we can immediately adapt the scheme to multiple simultaneous switching just by using bigger GMZIs. Finally, by a modification (an optical fibre connection instead of the entangling gate) we can implement the direct entangling protocol for the deterministic photon distribution in Fig.~\ref{fig:all2all} (right).

\subsection*{Case study: spin-photon entanglement}
\begin{figure}[b]
  \resizebox{4.5cm}{!}{\includegraphics{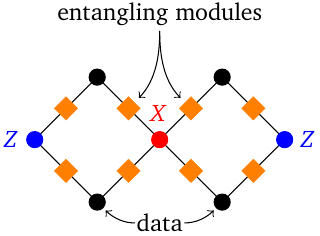}}
  \caption{$[4,1,2]$ surface code patch constructed from the T~centers~\cite{simmons2024scalable} serving as the data (black dots) and check qubits (red and blue dots) connected by the BSM entangling modules in the form of the Barret-Kok protocol~\cite{barrett2005efficient} depicted as orange squares.}
  \label{fig:prob412}
\end{figure}

\begin{figure}[t]
  \resizebox{13.5cm}{!}{\includegraphics{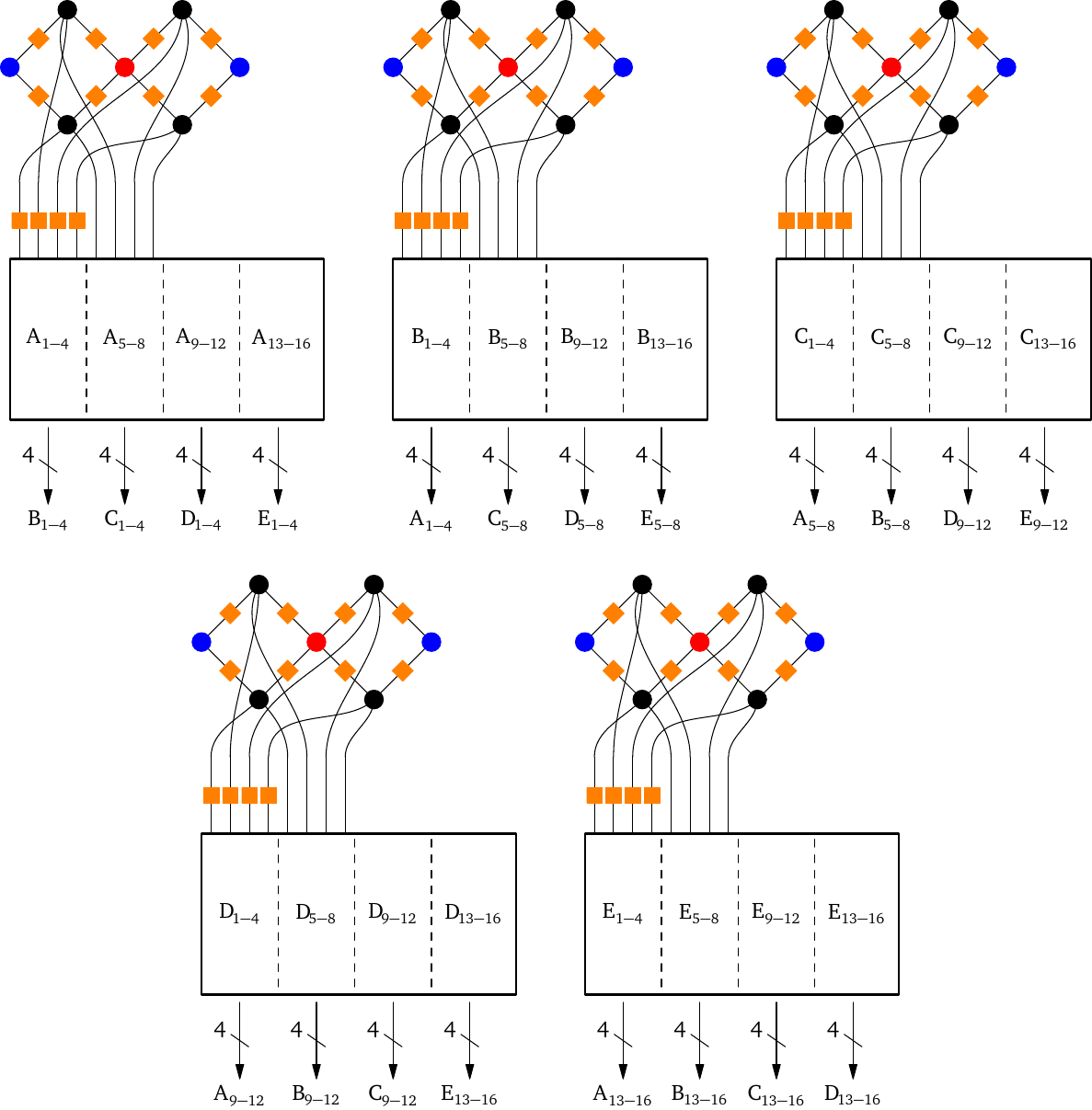}}
  \caption{A GMZI-based switching setup for a simultaneous transversal  CNOT between any pair of the five  $[4,1,2]$ surface code patches from Fig.~\ref{fig:prob412}. The patches are enhanced to enable the probabilistic CNOT (see the main text) and are connected to the hardwired $8\to16$ GMZIs labeled A,B,C,D and~E. Quadruples of input/output GMZI ports are labeled, e.g. A$_{1-4}$, to indicate the wiring address. The arrows point to where the photons go, not the directionality of how they travel in the fibers.}
  \label{fig:trans412}
\end{figure}

We will illustrate the use of the  GMZI-based switching schemes introduced in Sec.~\ref{sec:GMZIswitching} on the T~center/photon entanglement architecture~\cite{simmons2024scalable} as our first case study.  As before, our testing example will be a $[4,1,2]$ surface code patch and the whole setup is depicted in Fig.~\ref{fig:prob412}.  Both data and check qubits are the  T~centers and they emit photons entangled with the matter qubits which are electron spin states. The photons are probabilistically entangled in the Barret-Kok entangling gadget~\cite{barrett2005efficient} which is a specific implementation of the BSM. Its success probability is thus 50$\%$ in the lossless case and the resulting gate is essentially a CNOT gate between two remote matter qubits. If the CNOT gate succeeds the remote electron spins get entangled and it is transferred to the nuclear spins of the T~centers serving as a quantum memory. The failure to entangle is heralded in which case the entanglement  attempt is repeated~\cite{lim2006repeat}. Note that the photon-mediated entangling scheme  can be readily adapted (at least in principle) to a wide array of other color center quantum computing  platforms~\cite{nemoto2014photonic,knaut2024entanglement,bhaskar2020experimental} as well as other spin-photon physical system such as quantum dot emitters~\cite{de2024spin}. The long-lived nuclear quantum memory may, however, not always be  available (at all or with inferior parameters).

\begin{center}
    \renewcommand{\arraystretch}{1.2}
    \extrarowheight=\aboverulesep
    %\addtolength{\extrarowheight}{\belowrulesep}
    \aboverulesep=0pt
    \belowrulesep=0pt
    \begin{table}[h]
           \begin{tabular}{@{}>{\columncolor{white}[0pt][\tabcolsep]}  *{3}c @{}}
           \toprule
            {\cellcolor{lightgray}\ Paired modules } &   $\bphi$    & $\s$   \\
                        \midrule
             %A
             {\cellcolor{lightgray}\ A}  &  $(0000\,0000\,0000\,0000)$  &   $()$   \\
             {\cellcolor{lightgray}\ B}  &  $(\pi\pi\pi\pi\,0000\,\pi\pi\pi\pi\,0000)$  &   $\cdots(1,5)(2,6)(3,7)(4,8)\cdots$   \\
             \hline
             {\cellcolor{lightgray}\ A} &  $(\pi\pi\pi\pi\,0000\,\pi\pi\pi\pi\,0000)$  &   $\cdots(1,5)(2,6)(3,7)(4,8)\cdots$   \\
             {\cellcolor{lightgray}\ C} &  $(\pi\pi\pi\pi\,0000\,\pi\pi\pi\pi\,0000)$  &   $\cdots(1,5)(2,6)(3,7)(4,8)\cdots$    \\
             \hline
             {\cellcolor{lightgray}\ A} &   $(\pi\pi\pi\pi\,\pi\pi\pi\pi\,0000\,0000)$  &   $\cdots(1,9)(2,10)(3,11)(4,12)\cdots$  \\
             {\cellcolor{lightgray}\ D}   &  $(\pi\pi\pi\pi\,0000\,\pi\pi\pi\pi\,0000)$ &   $\cdots(1,5)(2,6)(3,7)(4,8)\cdots$  \\
             \hline
             {\cellcolor{lightgray}\ A} &   $(0000\,\pi\pi\pi\pi\,\pi\pi\pi\pi\,0000)$ &   $\cdots(1,13)(2,14)(3,15)(4,16)\cdots$    \\
             {\cellcolor{lightgray}\ E}   &  $(\pi\pi\pi\pi\,0000\,\pi\pi\pi\pi\,0000)$ &   $\cdots(1,5)(2,6)(3,7)(4,8)\cdots$   \\
             \hline
             %B
             {\cellcolor{lightgray}\ B}  &  $(\pi\pi\pi\pi\,0000\,\pi\pi\pi\pi\,0000)$  &   $\cdots(1,5)(2,6)(3,7)(4,8)\cdots$  \\
             {\cellcolor{lightgray}\ C}  &  $(0000\,0000\,0000\,0000)$  &   $()$   \\
             \hline
             {\cellcolor{lightgray}\ B}  &   $(\pi\pi\pi\pi\,\pi\pi\pi\pi\,0000\,0000)$  &   $\cdots(1,9)(2,10)(3,11)(4,12)\cdots$   \\
             {\cellcolor{lightgray}\ D}  &  $(0000\,0000\,0000\,0000)$  &  $()$   \\
             \hline
             {\cellcolor{lightgray}\ B}   &   $(0000\,\pi\pi\pi\pi\,\pi\pi\pi\pi\,0000)$ &    $\cdots(1,13)(2,14)(3,15)(4,16)\cdots$   \\
             {\cellcolor{lightgray}\ E}  &  $(0000\,0000\,0000\,0000$  &   $()$   \\
             \hline
             %C
             {\cellcolor{lightgray}\ C} &   $(\pi\pi\pi\pi\,\pi\pi\pi\pi\,0000\,0000)$  &  $\cdots(1,9)(2,10)(3,11)(4,12)\cdots$  \\
             {\cellcolor{lightgray}\ D}   & $(0000\,\pi\pi\pi\pi\,\pi\pi\pi\pi\,0000)$  &   $\cdots(5,9)(6,10)(7,11)(8,12)\cdots$  \\
             \hline
             {\cellcolor{lightgray}\ C}  &   $(\pi\pi\pi\pi\,\pi\pi\pi\pi\,0000\,0000)$  &  $\cdots(1,9)(2,10)(3,11)(4,12)\cdots$ \\
             {\cellcolor{lightgray}\ E}   & $(0000\,\pi\pi\pi\pi\,\pi\pi\pi\pi\,0000)$   &  $\cdots(5,9)(6,10)(7,11)(8,12)\cdots$\\
             \hline
             %D
             {\cellcolor{lightgray}\ D}   &   $(0000\,\pi\pi\pi\pi\,\pi\pi\pi\pi\,0000)$ &   $\cdots(1,13)(2,14)(3,15)(4,16)\cdots$  \\
             {\cellcolor{lightgray}\ E}   &   $(0000\,\pi\pi\pi\pi\,\pi\pi\pi\pi\,0000)$  &   $\cdots(1,13)(2,14)(3,15)(4,16)\cdots$ \\
             \hline
             \bottomrule
             \hline
            \end{tabular}\\ \vskip .3cm
    \caption{$8\to16$ GMZI phase configurations $\bphi$ and the resulting permutations $\s$ necessary to implement the connectivity depicted in Fig.~\ref{fig:trans412}. $\s$ is expressed in terms of a product of 2-cycles and we show only the relevant ones (see a full example in Eq.~\eqref{eqs:all2cycles}).}
    \label{table:N16example}
    \end{table}
\end{center}

Similar to the direct scenario from the previous section we consider $N=5$ remote $[4,1,2]$ surface code patches as our logical modules. The goal is to implement a  two-qubit logical Clifford gate (a logical CNOT in particular) in a fault-tolerant manner. This will be done by a transversal CNOT and a lattice surgery-based CNOT~\cite{horsman2012surface} (namely the merge operation as its crucial step). How does the GMZI connectivity changes compared to the direct entangling method in Fig.~\ref{fig:412direct}? The GMZI switching analysis of how to set up the switching phase shift angles developed in Appendix~\ref{sec:GMZImath} is again a crucial component in making the switching scheme work. As we indicate in Fig.~\ref{fig:trans412}, following the rough outline of Fig.~\ref{fig:GMZIswitch4}, we place the entangling gadgets between a code patch and a GMZI so that we don't artificially split the modules into a group of senders and receivers. The solution is, therefore, to double the number of the GMZI input ports so that one quadruple of inputs is connected to the entanglers and the other quadruple is not. We can then achieve sa2a by choosing any pair of modules such that only one of them uses the BSM entanglers. Since we route one of two quadruples of input states to five destinations we need to use an $8\to16$ GMZI per module and appropriately hardwire their output ports.

Let's  show the full connectivity of Fig.~\ref{fig:trans412} in Table~\ref{table:N16example} of how to set the phases for all 10 possible connected pairs of modules and the corresponding permutations~$\s$. The permutations are products of 16 two-cycles and they all do not fit in any reasonable way to the table so we write down just the four relevant ones for the linked module pairs. But  to have at least one complete derivation following the machinery developed in Appendix~\ref{sec:GMZImath} we show the following example useful, among others, for the A module of the~AE pairing: Let
$$
\bphi= (0000\,\pi\pi\pi\pi\,\pi\pi\pi\pi\,0000)
$$
from which we read off the permutations~$\s$:
\begin{align}\label{eqs:all2cycles}
  \s &=(1,9)(2,10)(3,11)(4,12)(5,13)(6,14)(7,15)(8,16)\nn\\
  &\times(1,5)(2,6)(3,7)(4,8)(9,13)(10,14)(11,15)(12,16)\nn\\
  &=(1,13)(5,9)(2,14)(6,10)(3,15)(7,11)(4,16)(8,12)\nn\\
  &=(1,13)(2,14)(3,15)(4,16)(5,9)(6,10)(7,11)(8,12),
\end{align}
where the third row is a reduction of a product of four two-cycles to two two-cycles. For instance, one of them is
$$
(1,9)(5,13)(1,5)(9,13)=(1,13)(5,9),
$$
see Eqs.~\eqref{eqs:productOfFour} for a derivation. The last row is obtained by reordering the commuting two-cycles and the leftmost product of four  is shown in Table~\ref{table:N16example} (the A row of the AE cell, for example).

Two comments are in place. Table~\ref{table:N16example} indicates one of four possible phase configurations for each pair of modules. The redundancy comes from the choice to use the BSM attached to one or the other module in the pair. Furthermore, we opted for the phase configurations whose overall phase is positive. But there is an equal number of the same switching configurations with a negative phase. Our second comment is about the hardwiring order of the GMZI output ports. We could have chosen to hardwire the modules differently, resulting in different phase shifts but achieving the same routing functionality. Even the order of each quadruple of fibers connecting the pairs of modules has some freedom in the way it could be wired (although it is far from the full freedom of all $4!$ permutations for each quadruple). It could be interesting to explore whether this redundancy has a potential to increase the switching capabilities of the GMZIs even further.

\begin{figure}[t]
  \resizebox{11cm}{!}{\includegraphics{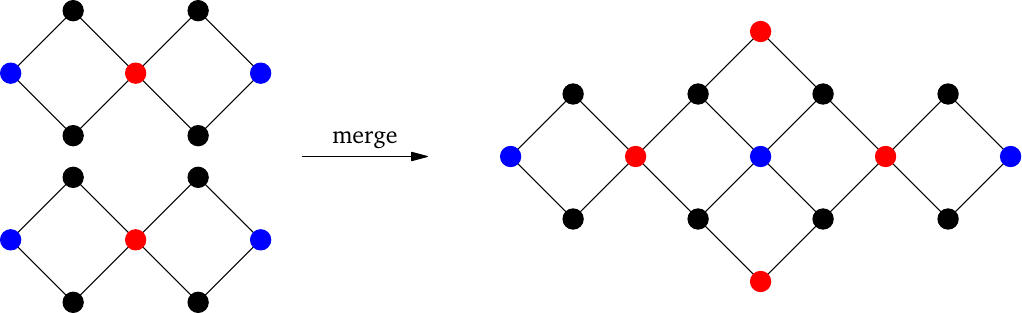}}
  \caption{Two  $[4,1,2]$ surface code patches combined (merged) into a single code.}
  \label{fig:mergePCM}
\end{figure}

How does the GMZIs achieve sa2a for logical CNOT using the merge lattice surgery operation?  Staying with the $[4,1,2]$ example, the merge operation in the language of parity check matrices (PCMs) is basically a creation of a bigger PCM, which can be conveniently visualised using the corresponding CSS code's Tanner graphs, see Fig.~\ref{fig:mergePCM}. The goal of the merge operation is to be able to measure a new set of stabilizers corresponding to a new (merged) code and so the role of the GMZI is to enable this measurement by connecting the corresponding data and check qubits.

\begin{figure}[t]
  \resizebox{14.5cm}{!}{\includegraphics{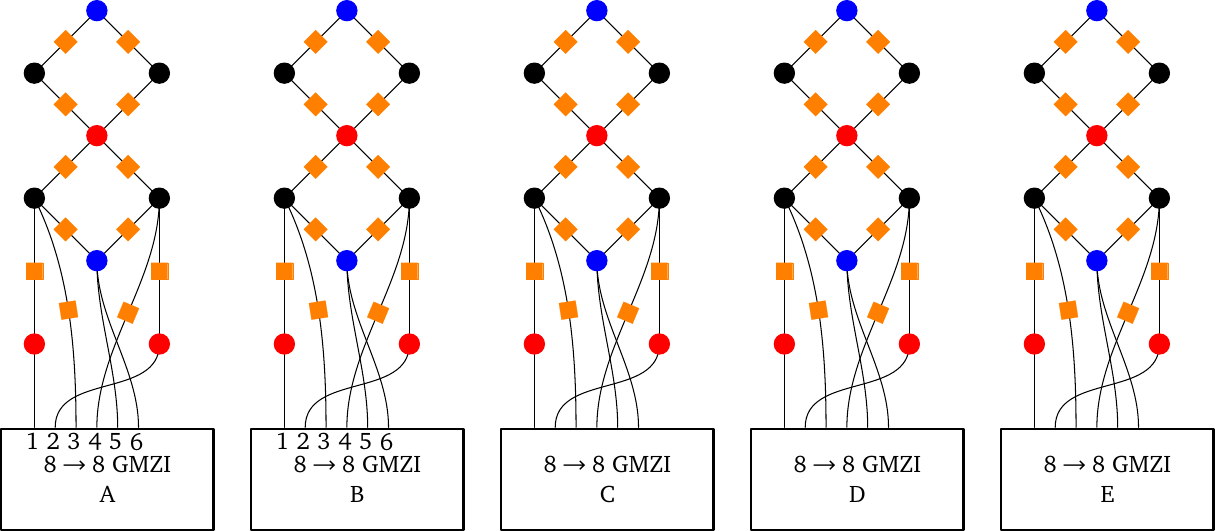}}
  \caption{Five $[4,1,2]$ modules (A-E) from Fig.~\ref{fig:prob412} enhanced so that they can be used to implement the merge operation in Fig.~\ref{fig:mergePCM}. They are hardwired such that the  switching capabilities of the $8\to8$ GMZI enable an sa2a merge operation between any pairs of modules.}
  \label{fig:412mergeGMZI}
\end{figure}
To this end, all modules carrying the surface code $[4,1,2]$ will be equipped with two additional $X$ check qubits (each is a T~center connected to a data qubit T~center in the same logical module via a BSM gadget) but also with an additional edge (that is, an additional BSM gadget) connected to the same data qubit, see Fig.~\ref{fig:412mergeGMZI}. A slightly different strategy is employed for the merge operation involving one of the $Z$~checks, where for that purpose we instead recycle one of the boundary $Z$~checks to become the future $Z$~check (the `middle' one in the merged PCM in Fig.~\ref{fig:mergePCM}), see again Fig.~\ref{fig:412mergeGMZI}. The switching device every logical module is equipped with is the $8\to8$ GMZI that will make sa2a merge (hence between any pair of the $N=5$ patches) possible.

\begin{center}
    \renewcommand{\arraystretch}{1.2}
    \extrarowheight=\aboverulesep
    %\addtolength{\extrarowheight}{\belowrulesep}
    \aboverulesep=0pt
    \belowrulesep=0pt
    \begin{table}[h]
           \begin{tabular}{@{}>{\columncolor{white}[0pt][\tabcolsep]}  *{3}c @{}}
           \toprule
            {\cellcolor{lightgray}\ Module/Stabilizer}  &   $\bphi$   & $\s$   \\
                        \midrule
             %A
             {\cellcolor{lightgray}\ A/$X$}  &  $(0000\,0000)$ &   $()$   \\
             {\cellcolor{lightgray}\ B/$X$}  &  $(\pi\pi00\,\pi\pi00)$  &   $(13)(24)(57)(68)$    \\
             \hline
             {\cellcolor{lightgray}\ A/$Z$}   &  $(\pi\pi\pi\pi\,0000)$ &   $(15)(26)(37)(48)$   \\
             {\cellcolor{lightgray}\ B/$Z$}  &  $(\pi\pi00\,\pi\pi00)$ &    $(13)(24)(57)(68)$   \\
             \hline
             \bottomrule
             \hline
            \end{tabular}\\ \vskip .3cm
    \caption{An illustration of the $8\to8$ GMZI connectivity needed for the merge stabilizer measurement for modules A and B in Fig.~\ref{fig:412mergeGMZI}.}
    \label{table:Mergeexample}
    \end{table}
\end{center}

As it turns out we need to route only two photons at a time to implement the desired stabilizer measurement. So the modules are hardwired by just two fibers. For example, \emph{output} ports A$_{12}$ are connected to B$_{12}$, A$_{34}$ to C$_{12}$ and so on all the way to D$_{78}$ to C$_{78}$. The logical modules are connected to the GMZI input ports $1$ to $6$ (as indicated  for GMZI A and B in Fig.~\ref{fig:412mergeGMZI}). Given the hardwiring, the GMZIs must provide the connectivity such that they either  links \emph{input} ports 1,2 with 3,4 of two different modules to update the new (merged) $X$ stabilizers or 3,4 to 5,6 for the merged $Z$ stabilizer\footnote{Input port 6 is, in fact, redundant since the $Z$ check connected to ports 5,6 can't serve both data qubits coming from 3,4 at the same time.}. We won't describe all possible switching configurations in favor of an example to pair AB for both $X,Z$ stabilizers, see Table~\ref{table:Mergeexample}.

\subsection*{Case study: GHZ measurement-based MBQC}

The use of GMZIs doesn't have to be limited to photon-matter physical platforms where the photons merely mediate the distribution of quantum information but the actual quantum computing substrate is the matter itself. GMZIs can find its use in purely linear-photonic approaches as well where the photons themselves carry quantum information~\cite{li2010fault,barrett2010fault,auger2018fault,bartolucci2023fusion}. One of the best-suited FT quantum architectures to be modularised is the design presented in~\cite{pankovich2024high}, which is the original proposal for linear-optical quantum computation with arbitrary error-correcting codes. The building blocks of any QEC code (foliated~\cite{bolt2016foliated,brown2020universal,sahay2023tailoring,paesani2023high} to make them suitable for the MBQC model~\cite{raussendorf2007topological}) are quantum parity check codes~\cite{bacon2006quantum} encoded Bell pairs~\cite{ewert2017ultrafast,lee2019fundamental}, which are `mere' bipartite states, and they are then measured by an elementary linear-optical measurement~\cite{pankovich2024flexible} to form the computational substrate in the form of the fault-tolerant QEC code. This is in contrast to a more traditional approach, where the resource states themselves are  $n$-partite un/encoded states~\cite{gimeno2015three,bartolucci2023fusion,song2024encoded} with the parameter~$n$ directly linked to the weight of the measured stabilizers. The flexibility of using Bell pairs is not only in the ability to implement any foliated code by merely redirecting more Bell pairs to the measurement area but also in the context of modular quantum computing. In fact, one can argue that this approach completely erases the difference between centralized and modular arrangements for fault-tolerant quantum computing while being realistic by sending and routing just bipartite states.

\begin{figure}[t]
  \resizebox{13cm}{!}{\includegraphics{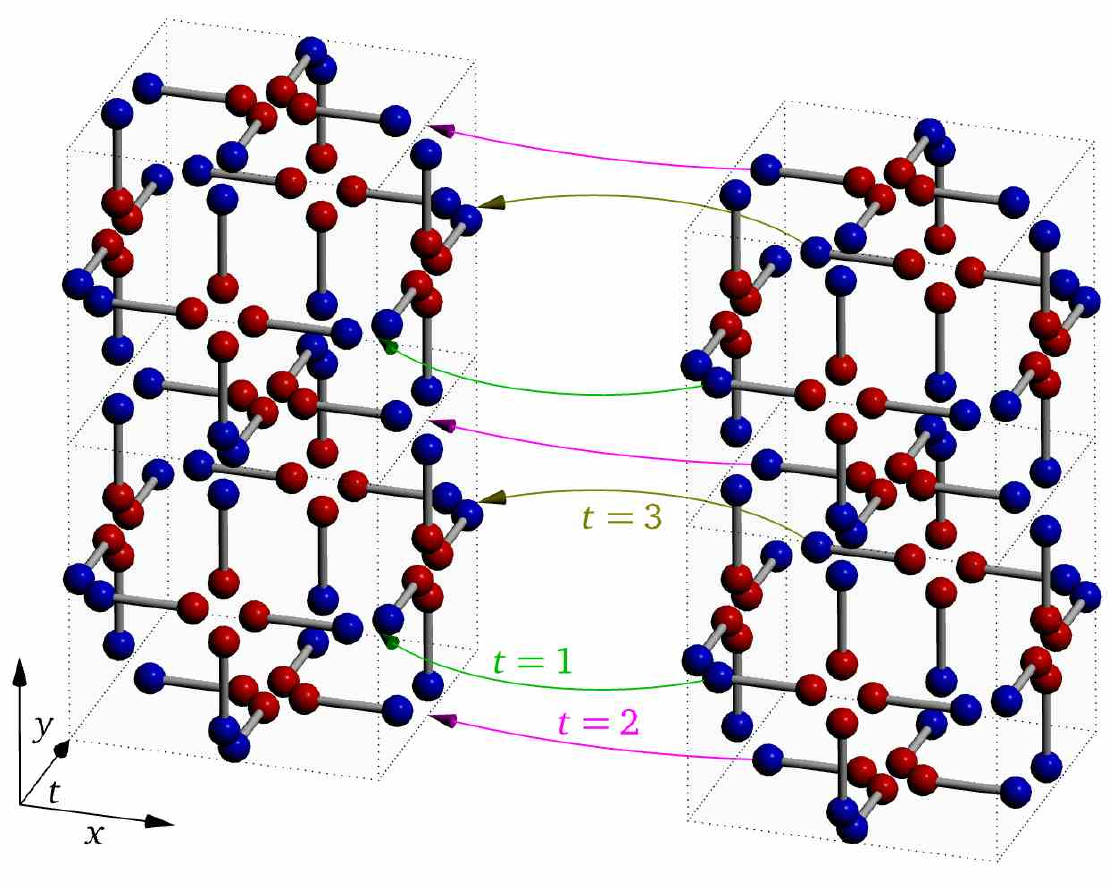}}
  \caption{RHG lattice~\cite{raussendorf2007topological} build from encoded Bell pairs (edges with a blue and a red vertex) and the GHZ measurement~\cite{pankovich2024high}. The arrows indicate how to redirect blue encoded qubits at different times from one idling substrate to the other in order to merge them. The RHG lattice is a foliated surface code but the GHZ measurement-based architecture works with any foliated CSS or stabilizer QEC code without a need to create other than encoded Bell pair graph states.}
  \label{fig:rhgghz}
\end{figure}

To illustrate the power of GMZIs as useful switching networks also in this scenario we won't attempt any sophisticated logical operations~\cite{herr2018lattice} -- our task is to enable such an operation among many modules via efficient switching. The architecture developed in~\cite{pankovich2024high} hasn't  been generalized in this direction even though the difficult, foundational, part in the form of the memory regime has been  established. We will instead simply consider two `columns' of the elementary RHG cells~\cite{raussendorf2007topological}  (the foliated surface code) whose edges are encoded Bell pairs and whose vertices are obtained by the GHZ measurement~\cite{pankovich2024high,pankovich2024flexible} and show how we would merge them in the spacelike direction using the switching techniques discussed here. An elementary operation of this type will undoubtedly figure out in the possible implementations of both Clifford and non-Clifford logical operations in these type of architecture.

To this end, one can envision a photonic architecture based on many different types of solid-state emitters (either probabilistic followed by multiplexing or deterministic) integrated on a single chip and serving as a resource state factory producing encoded photonic Bell states~\cite{tiurev2022high,lobl2024loss}. They are subsequently out-coupled and routed to the encoded Bell state measurement areas to implement any foliated CSS or stabilizer code. The resource state factory can be considered a module in the sense described in this paper. In that case, the switching setup is relatively simple.  In Fig.~\ref{fig:rhgghz} we see two separate idling vertically stacked RHG cells, where the blue/red edges are Bell pairs encoded in many variants of the QPC code~\cite{pankovich2024high,bacon2006quantum} (blue is intended for an encoded $Z$~stabilizer measurement and red for the $X$~version). There are three spacelike slices in different times $t=1,2,3$ and our goal is to merge both columns into a single idling substrate. This is the role of the GMZIs which will route the blue qubits forming Bell pairs from the left column as indicated by the arrows instead of routing them to the left wall of the right RHG column. So, for example, at $t=1$ (two green arrows) the $2\to4$ GMZI is used to route the blue qubits either to the same column (2 of the 4 GMZI outputs) or to the left column (the other two GMZI outputs) for measurement\footnote{Given an encoding of the blue qubit in $k$~photons the GMZI is going to be $2k\to4k$.}. The result is two merged columns of RHG cells (of course the extended stabilizer measurement on the previous boundaries must continue in the timelike direction). There may exist a huge variety of other possible scenarios for this and other photonic-based MBQC architectures, where the switching capabilities of the GMZI can be exploited.

\subsection*{Senders and receivers in quantum computing}
\begin{figure}[t]
  \resizebox{14cm}{!}{\includegraphics{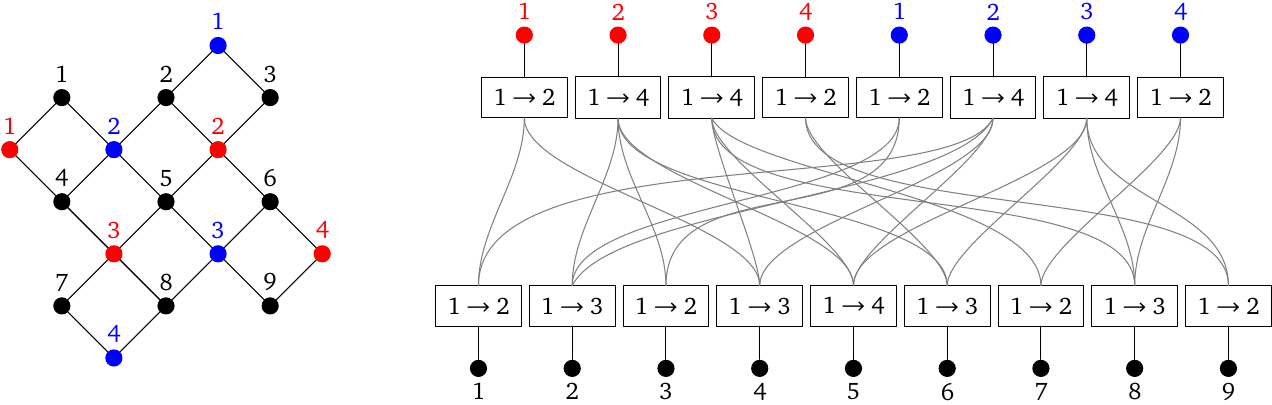}}
  \caption{The most straightforward use of the GMZIs is for a stabilizer measurement of the CSS codes. We depict  the $[9,1,3]$ rotated surface code's Tanner graph. On the left it is in a geometrically friendly form and on the right as a bipartite graph. The GMZIs depicted as rectangles are simple $1\to M$  switches. The main advantage of this scheme is a constant active depth two for any CSS code. The probabilistic entanglement strategy (to, for example, implement the $[4,1,2]$ stabilizer schedule in Fig.~\ref{fig:prob412}) would require an installation of entanglement modules on the check or data side of the graph (between a GMZI and a matter qubit).}
  \label{fig:stabdirect}
\end{figure}

Although sa2a is the main focus of this work, as our closing section we mention two situations where there is a natural split between senders and receivers in the quantum computing framework and the GMZIs naturally play a role: the stabilizer measurement of CSS codes and a distribution of magic states  to logical modules created in magic state factories. A stabilizer measurement for a CSS code can be visualized  with the help of its Tanner graph. This is a bipartite graph where the edges connect check and data qubits (the graph's vertices) as two disjoint sets of `senders' and `receivers'.  Stabilizer readout theory for CSS codes is full of sophisticated schemes where the ancillas are often prepared in highly entangled states~\cite{shor1996fault,steane1997active,knill2005quantum}. But the simplest and practically most often used technique is to reuse a single check qubit to conveniently measure a fixed number of neighboring data qubits -- precisely those indicated by the Tanner graph connectivity (for geometrically friendly Tanner graphs like the surface code). This leads to stabilizer readout schedules such as the well-known  fault-tolerant schedule \reflectbox{N},Z for the rotated surface code~\cite{tomita2014low}.

As an example we again consider the $[9,1,3]$ rotated surface code patch shown in Fig.~\ref{fig:stabdirect}. The GMZIs used in the way also depicted in Fig.~\ref{fig:stabdirect} enable any type of scheduling and the active depth is always two. The GMZIs are used as mere $1\to M$ or $N\to1$ switches. In fact, what we are looking at is nothing else than a custom-made Spanke's network. It has just enough sender-to-receiver connectivity to make sure all commuting stabilizer measurements can be done simultaneously.

Distributed magic state distillation factories is another example and in that case one can exploit additional favorable properties of the GMZIs we used in the sa2a case: the collective simultaneous routing of photons in many input ports. As before, assume a collection of modules supporting logical qubits. In this scenario, let there be a central magic state distillation (MSD) factory producing states useful to teleport non-Clifford logical operations in order to achieve universality. The MSD factories run fast enough to produce a necessary number of high-quality magic states that need to be routed to various logical modules. This is a suitable task for the GMZI since  thanks to our characterization results we know how to route a number $N$ of distilled magic states to $k$ different  locations with the help of the $N\to kN$ GMZI.

\section{Conclusions}

In this work we proposed several novel switching schemes for simultaneous any-to-any connectivity (sa2a) with the objective to enable inter-module connectivity of QEC code substrates supporting logical qubits. The main component of our design is the generalized Mach-Zehnder interferometer (GMZI), which is an optical component consisting of a layer of phase shifters sandwiched in between two passive linear-optical circuits. For specific phases, the GMZI acts as a limited $N\to M$ switch capable of routing any photon state from one or more of its input ports. As the state-of-the-art, a collection of $1\to M$ and $N\to 1$ GMZIs is used as part of  Spanke's switching network (hardwired on its output side), where it provides sa2a and has an active switching depth equal to four. Our presented GMZI-based switching designs outperform this already quite impressive design while achieving the same switching functionality. We were able to reduce the  active depth to two in addition to halving the number of chip-to-fiber couplers necessary for Spanke's network and for some variants also reduce the size or the number of necessary GMZI switches. The active depth and in- and out-couplers are significant sources of errors, namely photon loss, that needs to be minimized to enable scalable fault-tolerant quantum computing (FTQC). We  present  several GMZI-based proposals  that provide a photon-mediated matter qubit interaction as is required for distributed quantum computing platforms based on color centers, ions and neutral atoms  as well as for purely linear-photonic designs. Whether the photon/matter qubit interaction is deterministic or probabilistic, we are always guaranteed to beat the GMZI-based Spanke switch with our proposals.

The analysis was made possible by an exhaustive and purely quantum-mechanical analysis of a practically relevant class of $N\to M$ GMZIs, where $N,M$ are powers of two and whose passive optical circuit is a well-known $N$-dimensional quantum Fourier transform (QFT) circuit. Our method employs the powerful machinery of Wigner's $d$-matrices and we argue that this method is also suitable for future realistic description of the GMZIs, taking into account imperfect GMZI components or imperfect photon sources.

\appendix

\section{GMZIs and their mathematical properties}\label{sec:GMZImath}

In this technical appendix we study the quantum-mechanical behavior of the $N\to N$ GMZIs with no reference to its original description in terms of the (classical) transfer matrix mentioned in the main text. We choose $N=2^\lambda$ whose unitary MMIs will be denoted $W\in SU(N)$. It is known as Quantum Fourier Transform (QFT) and it can be conveniently generated by an iterative process detailed in~\cite{barak2007quantum} for any dimension $2^\lambda$. Thanks to this construction our results will also apply to a wider class of $N\to M$ GMZIs, where both $N$ and $M$ are powers of two. The $N$-dimensional QFT is implemented by a  passive optical circuit of depth $\log_2{N}=\lambda$ consisting of only $N/2\log_2{N}$ 50/50 (passive) beam splitters (BSs). The BS is the simplest `basic unit' that realizes an $SU(2)$ (or eventually $U(2)$) operation. The BSs act on  two-mode states which, depending on the number of photons, carry different representations of $SU(2)$. To properly describe their transformation we need to use the correct irreducible representation (irrep) of $SU(2)$. Since photons are bosons they live in the completely symmetric subspace of the Fock space (boson Fock space). The completely symmetric irrep  of $SU(2)$ is labelled by $j=1/2,1,\dots$, that is, the $j$-spin irrep, whose dimension is $2j+1$ with an orthogonal spanning basis $\ket{j,m}$ labeled by $m=-j,-j+1,\dots,+j$. We identify the total photon number~$n$ with $j=n/2$, so e.g., the multiplet subspace for $j=3/2$ is spanned by the bosonic Fock basis $\{\ket{n-\g,\g}\}_{\g=0}^3=\{\ket{3,0},\ket{1,2},\ket{2,1},\ket{0,3}\}$. Consequently, the action of a BS with $n$~input photons is given by~\cite{biedenharn1984angular}
\begin{equation}\label{eq:WignerMatrix}
   d^{(j)}(\vt)\df\exp{\big[-{\vt\over2}(J^{(j)}_+-J^{(j)}_-)\big]},
\end{equation}
where $J_\pm^{(j)}$ are the ladder operators of the spin-$j$ representation of $su(2)$ (the Lie algebra of $SU(2)$) of the following form:
\begin{equation}\label{eq:SU2stepOps}
  J_+^{(j)}=\sum_{m=-j}^{j-1}\sqrt{(j-m)(j+m+1)}\,\kbr{m-j+1}{m-j+2}=(J_-^{(j)})^\top,
\end{equation}
where $\top$ denotes the transpose. Conveniently, the $d^{(j)}(\vt)$ matrix, also known as the (small) Wigner matrix, has the following compact form
\begin{equation}\label{eq:dWignerwJacobi}
  d^{(j)}_{\mu,\nu}(\vt)
  =\left({(j+\mu)!(j-\mu)!\over(j+\nu)!(j-\nu)!}\right)^{1/2}\Big(\sin{\vt\over2}\Big)^{\mu-\nu}\Big(\cos{\vt\over2}\Big)^{\mu+\nu}
  P^{(\mu-\nu,\mu+\nu)}_{j-\mu}(\cos{\vt}),
\end{equation}
where $-j\leq\mu,\nu\leq j$ and $P^{(p,q)}_r(x)$ is the Jacobi polynomial
\begin{equation}\label{eq:JacobiP}
  P^{(p,q)}_r(x)=\sum_{s=0}^{r}{r+p\choose r-s}{r+q\choose s}\left({x-1\over2}\right)^s\left({x+1\over2}\right)^{r-s}.
\end{equation}
Note that the small Wigner matrix does not generate the whole group $SU(2)$ since we haven't incorporated the $J_z$ algebra generator anywhere (we don't need it for the special type of passive circuits we deal with in this work). That is the domain of the big Wigner matrix~\cite{biedenharn1984angular}.

It has been well-known since some time how the beam splitter acts on any input Fock state~\cite{kim2002entanglement}. But the Wigner matrix formalism is quite powerful in its compactness and versatility. By  setting $n=2j$ we will use the following notation $W_{k,l}(j,\vt)\df\big(d^{(j)}_{\mu,\nu}(\vt)\big)_{k,l}$ for the Wigner matrix describing the action of a beam-splitter acting on its input modes $k,l$ in a vector space spanned by the basis $\{\ket{n-\g,\g}_{k,l}\}$ for $0\leq\g\leq 2j$:
\begin{equation}\label{eq:dWignerAction}
  W_{k,l}(j,\vt)\ket{n-\g,\g}_{k,l}.
\end{equation}
The $N$-mode MMI unitary is of the form~\cite{barak2007quantum}
\begin{equation}\label{eq:MMIunitary}
W=\prod_{i=1}^{N/2\log_2{N}} W_{k_i,l_i}(j_i,\pi/2),
\end{equation}
where $d\equiv\dim{W}\leq{N+n_\text{tot}-1\choose n_\text{tot}}$ and $n_\text{tot}=\sum_{i=1}^{N}n_i$. The spanning Fock basis of $W$ is
\begin{equation}\label{eq:multiPhotonBasis}
  \big\{\ket{n^{(q)}_1,n^{(q)}_2,\dots,n^{(q)}_N}\big\}_{q=1}^d
\end{equation}
and $W\in\bbR^{d\times d}$ so $W^\dagger=W^\top$. The GMZI's unitary is then $W^\dagger DW$, where
\begin{equation}\label{eq:DphaseShifts}
  D(\bphi)=\diag{[e^{i\sum_{i=1}^{N}n^{(1)}_i\phi_i},\dots,e^{i\sum_{i=1}^{N}n^{(d)}_i\phi_i}]}
\end{equation}
is diagonal in basis~\eqref{eq:multiPhotonBasis} and $\bphi=(\phi_1,\dots,\phi_N)$ is an $N$-tuple of phase shift angles. We will show that the action of $\euS_{\bphi}=W^\dagger D(\bphi)W$ on a general input Fock state $\ket{n_1,n_2,\dots,n_N}$ for $n_i\geq0$ and carefully chosen phase angles $\bphi$ results in
\begin{equation}\label{eq:noInterference}
  \ket{n_1,\dots,n_N}\overset{\euS_{\bphi}}{\mapsto}\pm\ket{n_{\s^{-1}{(1)}},\dots,n_{\s^{-1}{(N)}}},
\end{equation}
where $\s\in S\subset S_N$ where $|S|\ll|S_N|$ for even modest $N$.

To this end, we start with a simple observation about the Jacobi function behavior. Let's set $x=0$ and swap $p$ and $q$ in~\eqref{eq:JacobiP}. Then
\begin{subequations}\label{eqs:JacobiPx0}
\begin{align}\label{eq:JacobiPx0a}
  P^{(q,p)}_r(0)&=\sum_{s=0}^{r}{r+q\choose r-s}{r+p\choose s}(-)^s\left({1\over2}\right)^{r}\\
                &=\sum_{s'=0}^{r}{r+q\choose s'}{r+p\choose r-s'}(-)^{r-s'}\left({1\over2}\right)^{r}\\
                &=(-)^r\sum_{s'=0}^{r}{r+p\choose r-s'}{r+q\choose s'}(-)^{s'}\left({1\over2}\right)^{r}\\
                &=(-)^r P^{(p,q)}_r(0)\label{eq:JacobiPx0d},
\end{align}
\end{subequations}
where the second row is obtained by setting $s'=r-s$ and in the third row we used $(-)^s=(-)^{-s}$. So we can swap the upper indices of the Jacobi function for $x=0$ and the function changes at most its sign. The consequence for the Wigner matrix entries from Eq.~\eqref{eq:dWignerwJacobi} is that for $\vt=\pi/2$ (because $x=\cos\vt$, cf.~\eqref{eq:dWignerwJacobi} and~\eqref{eq:JacobiP}) we can swap $\nu$ and $-\nu$ and the entry changes at most its sign: $d^{(j)}_{\mu,\nu}(\pi/2)=\pm d^{(j)}_{\mu,-\nu}(\pi/2)$. The rest of the Wigner matrix expression from~\eqref{eq:dWignerwJacobi} is invariant. The columns of both even- and odd-dimensional $W_{k,l}(n,\pi/2)$ are mirror-symmetric w.r.t the central vertical axis up to the overall matrix row sign whose origin lies in $(-)^r$ in~\eqref{eq:JacobiPx0d}. This factor alternates the overall row signs since $r=j-\mu$ (see~\eqref{eq:dWignerwJacobi}) and $\mu$ is the row matrix index.

\begin{figure}[t]
  \resizebox{14.5cm}{!}{\includegraphics{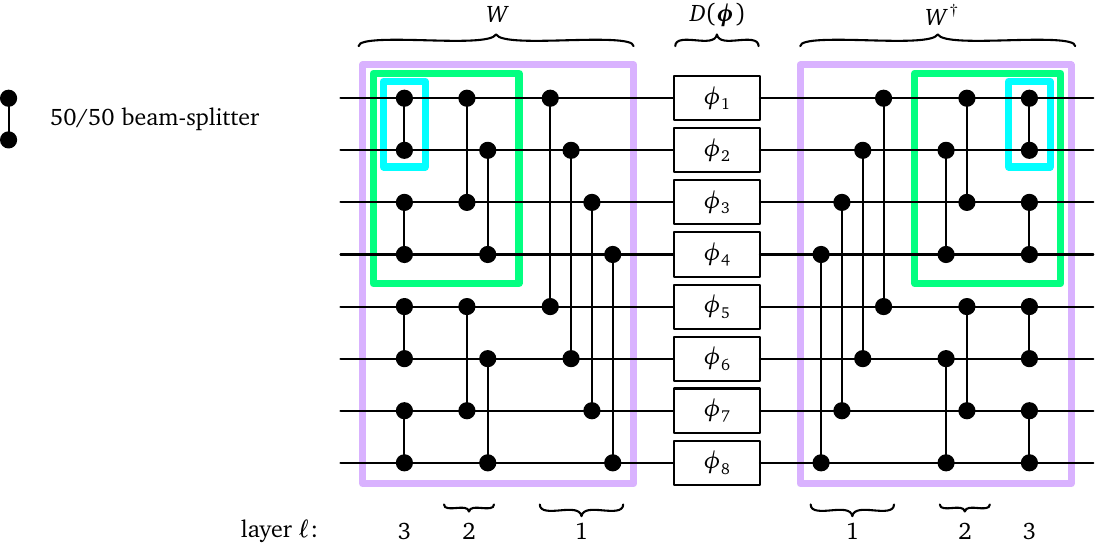}}
  \caption{The structure of the $8\to8$ GMZI and some terminology used to illustrate the derivation of the general $N\to N$ GMZI properties in order to act as an optical switch. $W$ is a passive optical circuit implementing the standard quantum Fourier transform~\cite{barak2007quantum} for a practically important case of $N$ being a power of two. The colored boxes indicate the iterative QFT construction. The operation $D(\bphi)$ acts by changing the phases $\phi_i$ which for the switching purposes take a value of zero or $\pi$. }
  \label{fig:unpick}
\end{figure}

By the time the input Fock state~\eqref{eq:noInterference} propagates to the layer of phase shifters $D(\bphi)$ in the middle of the GMZI it is a complicated superposition. But we know that once $D(\bphi)$ is applied the state encounters~$W^\dagger$, that is, the reverse-ordered and transposed product of the elementary Wigner matrices in~\eqref{eq:MMIunitary}. This structure is helpful. Thanks to~\eqref{eqs:JacobiPx0} we can start unpicking it from the middle of the GMZI and, crucially, without explicitly dealing with a complicated state vector. As illustrated in Fig.~\ref{fig:unpick} for the GMZI for $N=2^\lambda$ inputs by setting $\lambda=3$ we will count the layers~$\ell=1,2,\dots,\lambda$ of the mutually commuting BSs (that is, commuting within each layer) from the `inside' in both directions (in the $W$ and $W^\dagger$ `directions'). In particular, every pair of phase angles $(\phi_k,\phi_{N/2+k})$  of $D(\bphi)$ for $k=1,2,\dots, N/2$ will modify the output state of~$W$
\begin{equation}\label{eq:psiW}
  \psi_W\propto\sum_{q=1}^{d}a_{q}
  \ket{n^{(q)}_1,n^{(q)}_{N/2+1}}_{1,N/2+1}\dots\ket{n^{(q)}_k,n^{(q)}_{N/2+k}}_{k,N/2+k}\dots\ket{n^{(q)}_{N/2}n^{(q)}_N}_{N/2,N}
\end{equation}
according to Eq.~\eqref{eq:DphaseShifts}:
\begin{equation}\label{eq:phasehifts}
  a_{q}\ket{n^{(q)}_k,n^{(q)}_{N/2+k}}_{k,N/2+k}\mapsto
  a_{q}e^{i(n^{(q)}_k\phi_k+n^{(q)}_{N/2+k}\phi_{N/2+k})}\ket{n^{(q)}_k,n^{(q)}_{N/2+k}}_{k,N/2+k}.
\end{equation}
We proceed by rewriting how $D$ in~\eqref{eq:DphaseShifts} acts pairwise on modes $(k,N/2+k)$ as
\begin{equation}\label{eq:DphaseShiftsFactored}
  D(\bphi)=\prod_{k=1}^{N/2}D(\phi_k,\phi_{N/2+k})
  =\prod_{k=1}^{N/2}\diag{[e^{i(n^{(1)}_k\phi_k+n^{(1)}_{N/2+k}\phi_{N/2+k})},\dots,e^{i(n^{(d)}_k\phi_k+n^{(d)}_{N/2+k}\phi_{N/2+k})}]}.
\end{equation}

For the switching purposes we choose the angles $(\phi_k,\phi_{{N/2+k}})=\{(0,0),(0,\pi),(\pi,0),(\pi,\pi)\}$ from now on and study the behavior of $D(\phi_k,\phi_{N/2+k})$. It acts on  a  two-mode Hilbert subspace is spanned by $\{\ket{n^{(q)}_k,n^{(q)}_{N/2+k}}_{k,N/2+k}\}$ and this is an opportunity to simplify the notation. The superscript $q$ introduced in~\eqref{eq:multiPhotonBasis} to count the total Hilbert space basis is now superfluous so we drop it. We only need to know through which  $SU(2)$ representation $D(\phi_k,\phi_{N/2+k})$ acts and so we denote it $D_j(\phi_k,\phi_{N/2+k})$ which indicates the $j$-th representation of $SU(2)$ of dimension $2j+1$. We further denote the spanning basis of the carrying space  $\{\ket{n_k-\g,\g}_{k,N/2+k}\}_{\g=0}^{2j}$ (as in~\eqref{eq:dWignerAction}), and we set $n_k=n^{(q)}_k+n^{(q)}_{N/2+k}\equiv2j,\g=n^{(q)}_{N/2+k}$ for a chosen subset of $q$'s labeling the $2j+1$ suitable bases $\{\ket{n^{(q)}_k,n^{(q)}_{N/2+k}}_{k,N/2+k}\}$ from~\eqref{eq:psiW}. The representation space will then be labeled using the common multiplet notation $[2j_1+1]\ox\dots\ox[2j_{N/2}+1]$. We provide a further insight into the total Hilbert space structure where $\psi_W$ lives in Eq.~\eqref{eq:SU2TotCarSpace} when the permutation parity is investigated.

For any $j$ and the pairs of phases $(\phi_k,\phi_{N/2+k})=\{(0,0),(\pi,\pi)\}$ we get from~\eqref{eq:DphaseShiftsFactored}
\begin{equation}\label{eq:phaseshiftUnitaryOp1}
  D_j(\phi_k,\phi_{N/2+k})=(-)^{2j\phi_k/\pi}\id_{2j+1}
\end{equation}
and for $(\phi_k,\phi_{N/2+k})=\{(0,\pi),(\pi,0)\}$ it takes the following form
\begin{equation}\label{eq:phaseshiftUnitaryOp2}
  D_j(\phi_k,\phi_{N/2+k})=(-)^{2j\phi_{k}/\pi}\diag{[\underbrace{1,-1,1,-1,\dots,(-1)^{2j}}_{2j+1}]}.
\end{equation}
Note that the action of $D_j$ depends on the state it acts on but this is precisely taken into account  by the representation index~$j$. Table~\ref{table:DjAction} illustrates Eqs.~\eqref{eq:phaseshiftUnitaryOp1} and \eqref{eq:phaseshiftUnitaryOp2} on two generic examples.
\begin{center}
    \renewcommand{\arraystretch}{1.2}
    \extrarowheight=\aboverulesep
    %\addtolength{\extrarowheight}{\belowrulesep}
    \aboverulesep=0pt
    \belowrulesep=0pt
    \begin{figure}[h]
           \begin{tabular}{@{}>{\columncolor{white}[0pt][\tabcolsep]}  c|*5c }
%           & \multicolumn{4}{c}{$(\phi_1\phi_2\phi_3\phi_4\phi_5\phi_6\phi_7\phi_8$)}  \\
           \toprule
            {\ $(\phi_k,\phi_{N/2+k})$}   & $\ket{40}$ &  $\ket{31}$  &  $\ket{22}$ &  $\ket{13}$ & $\ket{04}$   \\
                        \midrule
             { $(0,0)$}                                    & +  &  + &  +   &+  & +  \\
             \rowcolor{tablerows}{$(0,\pi)$}                        & +  &  $-$ &  +   &$-$  & +  \\
             \rowcolor{tablerows}{$(\pi,0)$}                        & +  &  $-$ &  +   &$-$  & +  \\
             {$(\pi,\pi)$}                                & +  &  + &  +   &+  & +  \\
             \bottomrule
             \hline
            \end{tabular}
            \hspace{.666cm}
            \begin{tabular}{@{}>{\columncolor{white}[0pt][\tabcolsep]}  c|*4c }
           \toprule
            {\ $(\phi_k,\phi_{N/2+k})$}   & $\ket{30}$ &  $\ket{21}$  &  $\ket{12}$ &  $\ket{03}$    \\
                        \midrule
             { $(0,0)$}                                    & +  &  $+$ &  +   &+    \\
             \rowcolor{tablerows}{$(0,\pi)$}       & +  &  $-$ &  +   &$-$    \\
             \rowcolor{tablerows}{$(\pi,0)$}       & $-$  &  + &  $-$   &+    \\
             {$(\pi,\pi)$}                                & $-$  &  $-$ &  $-$   &$-$    \\
             \bottomrule
             \hline
            \end{tabular}\\ \vskip .3cm
    \caption{The action of the phase operator $D(\phi_k,\phi_{N/2+k})$ (c.f.~\eqref{eq:phaseshiftUnitaryOp1} and~\eqref{eq:phaseshiftUnitaryOp2}) on a two-mode state carrying the integer (left) and half-integer (right) irrep of $SU(2)$. The `mixed' phase angle rows are highlighted.}
    \label{table:DjAction}
    \end{figure}
\end{center}

Following Fig.~\ref{fig:unpick}, we see that the action in both layers~$\ell=1$ for $\{\phi_k,\phi_{N/2+k}\}=\{(0,0),(\pi,\pi)\}$ and any $k$ is captured by
\begin{equation}\label{eq:WdgDW0pipi}
  W^\dagger_{k,{N/2+k}}(j,\pi/2)D_j(\phi_k,\phi_{N/2+k})W_{k,{N/2+k}}(j,\pi/2)=(-)^{2j\phi_k/\pi}\id_{2j+1}
\end{equation}
since according to~\eqref{eq:phaseshiftUnitaryOp1} $D_j$ is proportional to an identity and  the BS Wigner matrix  defined above Eq.~\eqref{eq:dWignerAction} is unitary. In the second case, the alternating signs of~\eqref{eq:phaseshiftUnitaryOp2} cause changes to the signs of $W_{k,{N/2+k}}(j,\pi/2)$ and thus exactly reproducing the effect of swapping $p$ and $q$ in~\eqref{eqs:JacobiPx0}. Henceforth, the product $D_j(\phi_k,\phi_{N/2+k})W_{k,{N/2+k}}(j,\pi/2)$ effectively mirror-reflects all the columns of the matrix $W_{k,{N/2+k}}(j,\pi/2)$ up to a sign (for the whole column). But that means that the LHS of Eq.~\eqref{eq:WdgDW0pipi} becomes a row-reversed identity (an antidiagonal identity) for all $j$ and~$k$ that we will denote~$\di_{2j+1}$:
\begin{equation}\label{eq:WdgDWpipi0}
  W^\dagger_{k,{N/2+k}}(j,\pi/2)D_j(\phi_k,\phi_{N/2+k})W_{k,{N/2+k}}(j,\pi/2)=(-)^{2j\phi_{k}/\pi}\di_{2j+1}.
\end{equation}

We now  see what our choice of phase shifts does: it either preserves or swaps the basis of the state living in the corresponding two-mode Fock subspace with an additional change of its signs. But so far we analyzed just a single BS from layer $\ell=1$, say for $k=1$, while there are $N/2$ commuting BS gates in total (see Fig.~\ref{fig:unpick} for $N=8$). The question is then whether we can choose any phase shift pairs   $(\phi_k,\phi_{N/2+k})$ for $1<k\leq N/2$. It turns out that if we choose one angle pair from either $\{(0,0),(\pi,\pi)\}$ (let's call it the identity type) or $\{(0,\pi),(\pi,0)\}$ (the swap type)  for $k=1$ (or any~$k$) we have to keep choosing from the same type for all the remaining pairs~$(\phi_k,\phi_{N/2+k})$ such that $(\phi_k,\phi_{N/2^\ell+k})$ is again of the identity or swap type for  $1\leq k\leq N/2^\ell$ and~$1<\ell\leq\log_2{N}\equiv\lambda$. We will call such $N$ tuples of angles \emph{type-consistent}. A weaker notion of type consistency we will use is \emph{level-$\ell$ type-consistent} when we want to emphasize the condition's validity for  $(\phi_k,\phi_{N/2^\ell+k})$ but not necessarily for $l>\ell$. Note that if $(\phi_k,\phi_{N/2^\ell+k})$ fails the level-$\ell$ type-consistency for $\ell$ it fails it for all levels  $l>\ell$.

\begin{exa}
  Let $[N/2^\ell]$ denote $1,\dots,N/2^\ell$ and $\phi_{[N/2^\ell]}\df\phi_1\dots\phi_{N/2^\ell}$. For $N=8$ and $\ell=1$ the swap type choice can be, for example, $(\phi_{[4]}|\phi_{4+{[4]}})=(0\pi\pi\pi\,\pi000)$. But the type condition fails to be satisfied for $\ell=2$ since $(\phi_{[2]}|\phi_{2+{[2]}})=(0\pi\,\pi\pi)$ is not of a single type ($(\phi_1,\phi_{3})=(0,\pi)$ versus $(\phi_2,\phi_{4})=(\pi,\pi)$). This can be fixed in a number of ways, such as $(\phi_{[4]}|\phi_{4+{[4]}})=(0\pi\pi0\,\pi00\pi)$, $(00\pi\pi\,\pi\pi00)$ and so on, where for both $\ell=2$ and $3$ the types are now consistent.
\end{exa}
Let's see why type-consistent configurations are sufficient for the ability of the GMZI to function as a switch. Later we will show why it is also necessary. Once a type-consistent angle configuration is chosen, the string $(\phi_{[N/2^\ell]}|\phi_{N/2+{[N/2^\ell]}})$ allows us to deduce the permutation of the state if the GMZI contained just layers $\ell=1$. The permutation is obtained from $(\phi_{[N/2^\ell]}|\phi_{N/2+{[N/2^\ell]}})$ in the form of a 2-cycle $(k,N/2+k)$ if $(\phi_k,\phi_{N/2+k})$ is the swap type or a product of two 1-cycles $(k)(N/2+k)$ if $(\phi_k,\phi_{N/2+k})$ is the identity type. Since 1-cycles are permutation identities and 2-cycles are transpositions we always get a product of transpositions. Revisiting the type-consistent examples above, we get
\begin{align}
  (0\pi\pi0\,\pi00\pi) & \to (15)(26)(37)(48), \\
  (00\pi\pi\,\pi\pi00) & \to (15)(26)(37)(48).
\end{align}
The permutations for both configurations are so far identical since if there are only layers $\ell=1$ we essentially have four decoupled MZIs ($2\to2$ GMZI). Note that we haven't used the type-consistency condition yet. Also note that when dealing with transpositions in this paper we will be using the left-to-right convention, that is, $(ij)(ik)=(ijk)$.

The situation changes once we start adding consecutive  layers. The presence of layers $1<\ell\leq\lambda$ `activates' the 1- and 2-cycles obtained from $(\phi_k,\phi_{N/2^\ell+k})$. Continuing our previous examples, we get the desired permutations:
\begin{align}
  (0\pi\pi0\,\pi00\pi) & \to \underbrace{(15)(26)(37)(48)}_{\ell=1}\ \underbrace{(13)(24)(57)(68)}_{\ell=2}\ \underbrace{(12)(34)(56)(78)}_{\ell=3}\label{eq:ex1}, \\
  (00\pi\pi\,\pi\pi00) & \to (15)(26)(37)(48)\ (13)(24)(57)(68)\label{eq:ex2}.
\end{align}
The products look complicated but we will show that the type-consistent $N$-tuples always reduce to a product of exactly $N/2$ \emph{disjoint} transpositions unless the $N$-tuple is $(00\dots0)$ or $(\pi\pi\dots\pi)$ in which case there are no transpositions. It is precisely the sought after permutation~$\s$ in Eq.~\eqref{eq:noInterference} that directs an input GMZI state to its output and effectively realizes the corresponding switching task.

The necessity follows from the following argument. For the $N\to N$ GMZI, where $N=2^\lambda$, let $(\phi_{[N/2^\ell]}|\phi_{N/2+{[N/2^\ell]}})$ fail to be level-1 type-consistent. If we wanted to fix it and make it level-1 type-consistent (we don't but it will help us prove the necessity) we can keep w.l.o.g. the first half, $\phi_k$, and then there are only two ways to a type-consistent angle sequence: either $\phi_{N/2+[N/2^\ell]}$ becomes $\phi_{[N/2^\ell]}$ or its inverse (defined as $0\to\pi$ and $\pi\to0$) to be denoted $\overline{\phi}_{[N/2^\ell]}$. The latter is of interest for us. Considering  $(\phi_{[N/2^\ell]}|\phi_{N/2+{[N/2^\ell]}})$ and  $(\phi_{[N/2^\ell]}|\overline{\phi}_{[N/2^\ell]})$ as two strings of length~$N$ we record the positions where the strings differ. When we  convert $(\phi_{[N/2^\ell]}|\phi_{N/2+{[N/2^\ell]}})$ into a product of transpositions then it can be separated into two disjoint permutations $\s_1$ and $\s_2$ described by transpositions `made up' of the positions where the two strings coincide and differ, respectively. Calling these position strings  $\omega,\varpi$,  we have $\om\cup\varpi=[N]$ and $\om\cap\varpi=\emptyset$. But that means that such a GMZI is factorized into two disjoint optical circuits where the input state must transform as
\begin{equation}%\label{eq:}
  \ket{n_1,\dots,n_N}\mapsto\pm\ket{n_{\s_1^{-1}{(\om_1)}},\dots,n_{\s_1^{-1}{(\om_{|\om|})}}}\ket{n_{\s_2^{-1}{(\varpi_1)}},\dots,n_{\s_2^{-1}{(\varpi_{|\varpi|})}}},
\end{equation}
which even  in the best of cases, where the smaller circuits accidentally remain valid GMZIs, the RHS cannot realize all possible permutations the LHS is capable of.

It remains to show two properties of the type-consistent $N$-tuples:how to reduce the transposition products to $N/2$ disjoint transpositions and to generalize our analysis to include the permutation sign. We start with the former by elaborating on the previous paragraphs and realizing what the type consistency implies: an $N$-tuple is type-consistent only if it is of the form $(\phi_{[N/2^\ell]}|\phi_{[N/2^\ell]})$, which will be called the \emph{identity type sequence}, or $(\phi_{[N/2^\ell]}|\overline{\phi}_{[N/2^\ell]})$ (\emph{the swap type sequence}) for all~$\ell$. The second option gives rise to a product of non-trivial transpositions in the $\ell$-th layer\footnote{We will write some of the symbolic transpositions  as 2-cycles for clarity, that is, with a comma separating the elements.}:
\begin{equation}\label{eq:nontrivTrans}
  \prod_{\iota=0}^{2^{\ell-1}-1}\prod_{\ups=1}^{N/2^\ell}(\ups+\iota N/2^{\ell-1},\ups+\iota N/2^{\ell-1}+N/2^\ell)
\end{equation}
as witnessed in~\eqref{eq:ex1} for $N=8$ and $\ell=1,2,3$ and so the overall permutation reads
\begin{equation}\label{eq:nontrivTransAllEll}
 \prod_{j\geq1}\underbrace{\prod_{\iota=0}^{2^{\ell_j-1}-1}\prod_{\ups=1}^{N/2^{\ell_j}}(\ups+\iota N/2^{\ell_j-1},\ups+\iota N/2^{\ell_j-1}+N/2^{\ell_j})}_{\ell_j},
\end{equation}
where the potentially missing $\ell$'s are the layers with no transpositions (when $(\phi_{[N/2^{\ell}]}|\phi_{[N/2^{\ell}]})$ happens like the absent $\ell=3$ in~\eqref{eq:ex2} for $N=8$). Note that~$j$ indexes only the present $\ell's$ in Eq.~\eqref{eq:nontrivTransAllEll}. If only $\ell_1\neq0$ then~\eqref{eq:nontrivTransAllEll} is only a product of four transpositions and the claim follows. So let's assume the opposite and choose the $\ups=1,\iota=0$ transposition. From the $\ell_1$ product we get $(ab)\df(1,1+N/2^{\ell_1})$ and then, according to~\eqref{eq:nontrivTrans} there must exist two other transpositions in~\eqref{eq:nontrivTransAllEll}: $(ac)\df(1,1+N/2^{\ell_2})$ for $\ups=1,\iota=0$ and $(bd)\df(1+N/2^{\ell_1},1+N/2^{\ell_1}+N/2^{\ell_2})$ for $\ups=1,\iota=1$ from the $\ell_2$ product. From them we create third transposition, $(cd)=(1+N/2^{\ell_2},1+N/2^{\ell_2}+N/2^{\ell_1})$, and verify that it exists in the $\ell_1$ product by setting $\ups=1+N/2^{\ell_2}$ (which is in the product range of~\eqref{eq:nontrivTransAllEll} for $\ell_1$ thanks to $\ell_2>\ell_1$) and $\iota=0$.

All transpositions commute within every $\ell_1$ product and as along as we keep the left-to-right convention between the neighboring products $\ell_1$ and $\ell_{2}$  we can gather the four transpositions in this order and simplify:
%\begin{subequations}
\begin{align}\label{eqs:productOfFour}
  (ab)(cd)(ac)(bd)&=(ab)(cd)(ac)(cd)(cd)(bd)\nn\\
  &=(ab)(ad)(cd)(bd)\nn\\
  &=(dab)(cd)(bd)\nn\\
  &=(ad)(bd)(cd)(bd)\nn\\
  &=(ad)(bc)\nn\\
  &=(1,1+N/2^{\ell_2}+N/2^{\ell_1})(1+N/2^{\ell_2},1+N/2^{\ell_1}),
\end{align}
%\end{subequations}
where in the first row we inserted an identity $(cd)(cd)$, in the second row we simplified the following conjugation
$$
(cd)(ac)(cd)=(cda)(cd)=(dac)(cd)=(ad)
$$
and similarly in the fifth row. The rest is standard rules for the action from the left that we follow in this paper. We repeat the same derivation $N/4$ times (one factor of 2 appears in the denominator because each transposition contains two elements and the other factor of 2 comes from the fact that on the LHS of the first row of~\eqref{eqs:productOfFour} we always `consume' two transpositions of $\ell_1$ and $\ell_2$).

Hence, generalizing the last row of~\eqref{eqs:productOfFour}, we get from~\eqref{eq:nontrivTransAllEll}
\begin{subequations}\label{eqs:nontrivTransTwoEll}
\begin{align}
  &\underbrace{\prod_{\iota=0}^{2^{\ell_j-1}-1}\prod_{\ups=1}^{N/2^{\ell_j}}(\ups+\iota N/2^{\ell_j-1},\ups+\iota N/2^{\ell_j-1}+N/2^{\ell_j})}_{\ell_j}\nn\\
  \times&\underbrace{\prod_{\iota=0}^{2^{\ell_{j+1}-1}-1}\prod_{\ups=1}^{N/2^{\ell_{j+1}}}(\ups+\iota N/2^{\ell_{j+1}-1},\ups+\iota N/2^{\ell_{j+1}-1}+N/2^{\ell_{j+1}})}_{\ell_{j+1}}\\
  =&\prod_{n=1}^{N/4}(\ups_n,\ups_n+N/2^{\ell_{j+1}}+N/2^{\ell_j})(\ups_n+N/2^{\ell_{j+1}},\ups_n+N/2^{\ell_{j}}),\label{eqs:nontrivTransTwoEllB}
\end{align}
\end{subequations}
where  $n$ indexes the participating elements $\ups_n$, which is just a permutation of $[N/2]$. By the above procedure we converted a non-commuting product of  two products of $N/2$ transpositions to a product of different  $N/2$ transpositions that mutually commute. Now we iteratively repeat the same process for the remaining $\ell_{j+h},h>1$ products. Let's illustrate Eq.~\eqref{eqs:nontrivTransTwoEll}.
\begin{exa}
  We pick where we left~\eqref{eq:ex1} and follow the last two rows of~\eqref{eqs:productOfFour} or its general form~\eqref{eqs:nontrivTransTwoEllB}:
  \begin{align}\label{eq:ex1cont}
    (0\pi\pi0\,\pi00\pi) &\to  \underbrace{\textcolor{cyan}{(15)}(26)\textcolor{cyan}{(37)}(48)}_{\ell_1}\ \underbrace{\textcolor{cyan}{(13)}(24)\textcolor{cyan}{(57)}(68)}_{\ell_2}\ \underbrace{(12)(34)(56)(78)}_{\ell_3}\nn\\
    &= \textcolor{cyan}{(17)(35)}(28)(46)\,(12)(34)(56)(78)\nn\\
    &= (18)(27)(36)(45),
  \end{align}
  where we highlighted one step in blue. Similarly, we get
  \begin{align}
    (0\pi0\pi\,\pi0\pi0) &\to  \underbrace{\textcolor{cyan}{(15)(26)}(37)(48)}_{\ell_1}\ \underbrace{\textcolor{cyan}{(12)}(34)\textcolor{cyan}{(56)}(78)}_{\ell_2}\nn\\
    &= \textcolor{cyan}{(16)(25)}(38)(47).
  \end{align}
\end{exa}

A noteworthy corollary of~\eqref{eqs:nontrivTransTwoEll} is that all $\ell_j$ products commute as long as we keep the orders fixed within each product. This is a consequence of the commutativity of~\eqref{eqs:nontrivTransTwoEllB}: we can suitably permute the transpositions and reassemble the first row of~\eqref{eqs:nontrivTransTwoEll} in the opposite order. Therefore, each $N\to N$ GMZI can be realized in $(\log_2{N})!$ different ways generalizing the obvious possibility $W\to W^\dagger$. The new configurations may eventually lead to better manufacturing options by simplifying the photonic chip design.

So far we focused on the permutation aspect of the GMZI but in reality this device implements a signed permutation whose parity depends on the $SU(2)$ representation of the space the input state carries. We now address the sign's origin for the case of $n_i=0,1$ for any $N\to N$ GMZIs for $N=2^k,k\geq1$ before turning our attention to an unconstrained~$n_i$.

As a remainder, the signs come from Eqs.~\eqref{eq:phaseshiftUnitaryOp1} and~\eqref{eq:phaseshiftUnitaryOp2} describing how the GMZIs phase shifts act on two-mode boson states of the total photon number~$n=2j$ carrying the $j$-th representation of $SU(2)$. Table~\ref{table:DjAction} illustrates the $D_j$'s action on~$n$ even and odd. The subsequent conjugation by the Wigner matrices in Eqs.~\eqref{eq:WdgDW0pipi} and~\eqref{eq:WdgDWpipi0} reveals not only whether the basis order swaps but also the accompanying signs. Since the overall permutation $\s$ of any input state~\eqref{eq:noInterference} was just classified we may solely focus on the sign factor and ignore the permutation.

Let start with the simplest scenario where the input state of the $N\to N$ GMZI is the $N$-mode state $\ket{1,0,\dots,0}$. We are thus guaranteed that at each step of the state evolution every pair of modes contains at most one photon and in particular this is true for the state $\psi_W$ in~\eqref{eq:psiW}. This state must be of the form $\psi_W=\sum_{i=1}^{d}a_i\ket{0,\dots,0,1_i,0,\dots,0}$ and therefore two modes carry the $j=1/2$ representation of $SU(2)$ (the $[2]$ multiplet) while the rest is a product of trivial $j=0$ representations (the $[1]$ multiplet, see the next example for more details). Note that in this case $d=N$ saturates the boson Hilbert space dimension written below~Eq.~\eqref{eq:MMIunitary} since $n_{\text{tot}}=1$ and all $a_i\neq0$ due to the structure of $W$ not causing any destructive interference.  Following~\eqref{eq:phaseshiftUnitaryOp1} and~\eqref{eq:phaseshiftUnitaryOp2} we get for  $(\phi_k,\phi_{N/2+k})=\{(0,0),(\pi,\pi)\}$
\begin{subequations}\label{eqs:Donehalf00ORpipi}
\begin{align}
  D_{1/2}(\phi_k,\phi_{N/2+k}) &= \begin{bmatrix}
                      \pm1 & 0 \\
                      0 & \pm1 \\
                    \end{bmatrix},
  \\
  D_0(\phi_k,\phi_{N/2+k}) &= 1
\end{align}
\end{subequations}
and for  $(\phi_k,\phi_{N/2+k})=\{(0,\pi),(\pi,0)\}$
\begin{subequations}\label{eqs:Donehalf0piORpi0}
\begin{align}
  D_{1/2}(\phi_k,\phi_{N/2+k}) &= \begin{bmatrix}
                      \pm1 & 0 \\
                      0 & \mp1 \\
                    \end{bmatrix},
  \\
  D_0(\phi_k,\phi_{N/2+k}) &= 1,
\end{align}
\end{subequations}
where the upper sign is for the first angle pair. It is instructive to write down the conjugation output of~\eqref{eq:WdgDW0pipi} and~\eqref{eq:WdgDWpipi0} for $j=1/2$:
\begin{align}\label{eq:WembraceDonehalf1}
  W^\dagger_{k,{N/2+k}}(1/2,\pi/2)\begin{bmatrix}
                      \pm1 & 0 \\
                      0 & \pm1 \\
                    \end{bmatrix}W_{k,{N/2+k}}(1/2,\pi/2)
                    & =
                    \begin{bmatrix}
                      \pm1 & 0 \\
                      0 & \pm1 \\
                    \end{bmatrix},\\
  W^\dagger_{k,{N/2+k}}(1/2,\pi/2)\begin{bmatrix}\label{eq:WembraceDonehalf2}
                      \pm1 & 0 \\
                      0 & \mp1 \\
                    \end{bmatrix}W_{k,{N/2+k}}(1/2,\pi/2)
                    & =
                    \begin{bmatrix}
                      0 & \mp1 \\
                      \mp1 & 0 \\
                    \end{bmatrix}.
\end{align}
This is all we need to know to find the permutation parity for any GMZI type we investigate here. To this end, we introduce a shorthand notation summarizing the above matrix transformations, where we just keep the information about how the signs of the matrix entries change and not how they are permuted:
\begin{subequations}\label{eqs:simplifiedConjugation}
\begin{align}%\label{}
  (1,1) & \to (1,1), \\
  (1,-1) & \to (-1,-1), \\
  (-1,1) & \to (1,1), \\
  (-1,-1) & \to (-1,-1).
\end{align}
\end{subequations}
The crucial observation is that the second digit on the left is preserved on the right and also gets copied to the first digit ibid. So an identity type sequence $\bphi=(\phi_{[N/2^\ell]}|\phi_{[N/2^\ell]})$ is mapped to itself corresponding to the first and last row of~\eqref{eqs:simplifiedConjugation} whereas the swap type sequence $\bphi=(\phi_{[N/2^\ell]}|\overline{\phi}_{[N/2^\ell]})$ becomes $(\overline{\phi}_{[N/2^\ell]}|\overline{\phi}_{[N/2^\ell]})$ corresponding to the second and third row of~\eqref{eqs:simplifiedConjugation}. But that also means that as we conjugate by the beam splitters in layers $\ell=1,2,3,\dots$ we always preserve the second half of either the identity or the swap sequence as it gets copied to the other halves (recall that type consistency means it is either identity or swap type on all levels~$\ell$). Consequently, for the sequence $\bphi$  of length $N$ it is its last digit, $\bphi(N)$, that determines the sign of the permutation: $(-)^{\bphi(N)/\pi}$.

The previously described procedure might seem contrived but it is actually quite straightforward. Before proceeding, let's illustrate it on a example.
\begin{exa}
  Let $N=8$ and $\bphi=(\phi_{[4]}|\phi_{4+{[4]}})=(\pi00\pi\,0\pi\pi0)$ be a type-consistent sequence which is of the swap type for all $\ell=1,2,3$ (i.e., both $(\phi_{[2]}|\phi_{2+{[2]}})$ and $(\phi_{[1]}|\phi_{1+{[1]}})$ are swap types). Note that this sequence is a negation of the one investigated in Eq.~\eqref{eq:ex1cont} so let's study both of them. We first construct $D$ according to~Eq.~\eqref{eq:DphaseShifts} and conjugate by the $W$'s for all layers~$\ell=1,2,3$ in this order. Given the form of $\psi_W=a_1\ket{1,0,\dots,0}+\dots+a_d\ket{0,\dots,0,1}$ for any input state $\ket{0,\dots,1_i,\dots,0}$  we can use the factorized form of~\eqref{eq:DphaseShifts}, Eq.~\eqref{eq:DphaseShiftsFactored}, to construct $D$, where $\psi_W$ carries the representation  ${1\over2}\ox0^{\ox 3}$ of $SU(2)$. We get from~\eqref{eqs:Donehalf0piORpi0}
\begin{align}\label{eq:Dforpi00pi0pipi0}
    D(\bphi) &= \bigoplus_{k=1}^4D_{1/2}(\phi_k,\phi_{4+k})\bigotimes_{\kappa\in[4]\backslash k}D_{0}(\phi_\kappa,\phi_{4+\kappa})
    \equiv\bigoplus_{k=1}^4D_{1/2}(\phi_k,\phi_{4+k})\nn\\
    &=\pm\diag{[-1,1,1,-1,1,-1,-1,1]}
\end{align}
for $\bphi=\{(\pi00\pi\,0\pi\pi0),(0\pi\pi0\,\pi00\pi)\}$.  If we wanted to find $W^\dagger DW$ the hard way we could repeatedly use the matrix representation from~\eqref{eq:WembraceDonehalf2}. We would get $W^\dagger DW=\pm\di_8$, which corresponds to the signed permutation $\s=(18)(27)(36)(45)$. But we already know how to efficiently obtain the permutation from the previous analysis and now we are interested just in the permutation sign. So instead we use the update rules from~\eqref{eqs:simplifiedConjugation} leading to
  \begin{align}\label{eqs:UpdateExample}
    \pm\diag{[-1,1,1,-1,1,-1,-1,1]} & \overset{\ell=1}{\to} \pm[1,-1,-1,1,1,-1,-1,1] \nn \\
      & \overset{\ell=2}{\to}  \pm[-1,1,-1,1,-1,1,-1,1]   \nn \\
      & \overset{\ell=3}{\to}  \pm[1,1,1,1,1,1,1,1]\nn\\
      & \equiv(-)^{\bphi(N)/\pi}[1,1,1,1,1,1,1,1],
  \end{align}
where the arrows indicate the  action of a quadruple of BSs from layer~$\ell$ (see Fig.~\ref{fig:unpick}). In each row of~\eqref{eqs:UpdateExample} we applied the update rules to the pairs of interacting modes and recovered the sign. We confirm in the last row that  indeed the sign is determined by the last element of~$\bphi$. The resulting input/output switching transformation is therefore
  $$
  \ket{0,\dots,1_i,\dots,0}\overset{\euS_{\bphi}}{\to}(-)^{\bphi(N)/\pi}\ket{0,\dots,1_{\s^{-1}(i)},\dots,0}
  $$
  where the information about both the sign and the permutation $\s$ is encoded in $\bphi$ as expected.
\end{exa}
We know from our previous permutation analysis, where we disregarded the sign, that the switch $\euS$ is oblivious to what the input state is. Clearly, the signed permutation for a given switching configuration is also input state-independent if exactly one single photon enters through any port. But that cannot be the case when we consider more single photons in the input modes. So how does the phase behave in that case? Considering the switch behavior  we get
\begin{align}\label{eqs:switchAction}
  \euS_{\bphi}(\ket{0,\dots,1_i,\dots,1_j,\dots,0}) &= \euS_{\bphi}(\ket{0,\dots,1_i,\dots,0})\ox\euS_{\bphi}(\ket{0,\dots,1_j,\dots,0}) \nn\\
   &= (-)^{\bphi(N)/\pi}\ket{0,\dots,1_{\s^{-1}(i)},\dots,0}\ox(-)^{\bphi(N)/\pi}\ket{0,\dots,1_{\s^{-1}(j)},\dots,0} \nn\\
   &= \ket{0,\dots,1_{\s^{-1}(i)},\dots,1_{\s^{-1}(j)},\dots,0}.
\end{align}
Hence, we may conclude that for $n_i=0,1$ and any $N$ we get
\begin{equation}\label{eq:inputStateTrans}
   \ket{n_1,\dots,n_N}\overset{\euS_{\bphi}}{\mapsto}(-)^{n_{\text{tot}}\bphi(N)/\pi}\ket{n_{\s^{-1}{(1)}},\dots,n_{\s^{-1}{(N)}}}.
\end{equation}

We now generalize~\eqref{eqs:switchAction} to the case of arbitrary $n_i\geq0$, that is, for any input state with the total photon number $n_{\text{tot}}$. To this end, we first assume the input state to be of the form $\ket{0,\dots,n_i,\dots,0}$, where $n_i=n_{\text{tot}}$ (all input photons in a single mode). Given the lack of destructive interference, the state $\psi_W$ explores the entire available Hilbert space of dimension $\binom{N+n_{\text{tot}}-1}{n_{\text{tot}}}$. The state carries the reducible representation of $SU(2)$ acting on
\begin{equation}\label{eq:SU2TotCarSpace}
  \psi_W\in\Hcal\simeq\bigoplus_{(j_1,\dots,j_{N/2})}\bigotimes_{i=1}^{N/2}\Hcal_{j_i},
\end{equation}
where $\Hcal_{j_i}$ corresponds to the $j_i$-th representation of $SU(2)$ and the sum goes over all compositions $\sum_{i=1}^{N/2}j_i=j\equiv n_{\text{tot}}/2$ for all integers and half-integers $n_{\text{tot}}/2\geq j_i\geq0$. Let's demonstrate this useful Hilbert space decomposition.
\begin{exa}
    Let $n_{\text{tot}}=3$ and $N=8$. Then
\begin{equation}\label{eq:HdecompositionEx}
    \Hcal\simeq\wp[\Hcal_{3/2}\ox\Hcal_{0}^{\ox3}]\oplus\wp[\Hcal_{1}\ox\Hcal_{1/2}\ox\Hcal_{0}^{\ox2}]\oplus\wp[\Hcal_{1/2}\ox\Hcal_{1/2}\ox\Hcal_{1/2}\ox\Hcal_{0}],
\end{equation}
where $\wp$ denotes the symmetrizer. Since $\dim{\Hcal_{j}}=2j+1$ we deduce $\dim{\Hcal}=\binom{N+n_{\text{tot}}-1}{n_{\text{tot}}}=\binom{10}{3}=120$ which agrees with $4\times4+6\times12+8\times4$  obtained from decomposition~\eqref{eq:HdecompositionEx}.
\end{exa}
We shall call the highest-weight Hilbert space sector $\Hcal_{n_{\text{tot}}/2}\ox\Hcal_0^{\ox(N-2)/2}\simeq\Hcal_{n_{\text{tot}}/2}$ (and any of its permutations, see the first direct summand on the RHS of~\eqref{eq:HdecompositionEx}) the \emph{pivot} and use it to deduce the permutation sign. The advantage of the pivot is its simple structure because we can ignore the trivial one-dimensional Hilbert space(s) $\Hcal_0$ spanned by a (two-mode) vacuum. The pivot Hilbert space $\Hcal_{n_{\text{tot}}/2}$ is spanned by the previously introduced basis $\{\ket{n-\g,\g}_{k,N/2+k}\}_{\g=0}^{2j}$ for $j=n_{\text{tot}}/2$.% and the other subspace forming $\Hcal$ the $SU(2)$ acts irreducibly according to the $j$-th representation.

To proceed, in a manner similar to the $n_{\text{tot}}=1$ case in~Eqs.~\eqref{eq:WembraceDonehalf1} and~\eqref{eq:WembraceDonehalf2}, we expand, for convenience, Eqs.~\eqref{eq:WdgDW0pipi} and~\eqref{eq:WdgDWpipi0}. We consider separately $n_{\text{tot}}$ even (for~$j$ integers) and  $n_{\text{tot}}$ odd (for~$j$ half-integers). For the former we get
\begin{subequations}\label{eq:WembraceDjf1}
\begin{align}%\label{eqs:}
   W^\dagger_{k,{N/2+k}}(j,\pi/2)\id_{2j+1}W_{k,{N/2+k}}(j,\pi/2)&=\id_{2j+1}, \\
   W^\dagger_{k,{N/2+k}}(j,\pi/2)\,\diag{[1,-1,1,-1,\dots,1]}\,W_{k,{N/2+k}}(j,\pi/2)&=\di_{2j+1}
\end{align}
\end{subequations}
for $(\phi_k,\phi_{N/2+k})=\{(0,0),(\pi,\pi)\}$ and $\{(0,\pi),(\pi,0)\}$. In the half-integer case we get
\begin{subequations}\label{eq:WembraceDjf2}
\begin{align}%\label{eqs:}
    W^\dagger_{k,{N/2+k}}(j,\pi/2)\,(\pm\id_{2j+1})\,W_{k,{N/2+k}}(j,\pi/2)&=\pm\id_{2j+1}, \\
    W^\dagger_{k,{N/2+k}}\,(j,\pi/2)(\pm\diag{[1,-1,1,-1,\dots,-1]})\,W_{k,{N/2+k}}(j,\pi/2)&=\mp\di_{2j+1}
\end{align}
\end{subequations}
again for \ $(\phi_k,\phi_{N/2+k})=\{(0,0),(\pi,\pi)\}$ and $\{(0,\pi),(\pi,0)\}$ in this order. Although the first rows of Eqs.~\eqref{eq:WembraceDjf1} and~\eqref{eq:WembraceDjf2} are trivial, all the expressions taken together reveal the sought-after general sign behavior. In the integer case, where $n_{\text{tot}}$ is even, the sign is always positive. This is because the RHS of~\eqref{eq:WembraceDjf1} is positive and this behavior is intuitively expected for even total photon numbers. For $n_{\text{tot}}$ odd we find exactly the same behavior as in~\eqref{eqs:simplifiedConjugation} by observing the two-dimensional subspace of the pivot space $\Hcal_{n_{\text{tot}}/2}$ spanned by $\{\ket{n_{\text{tot}},0},\ket{0,n_{\text{tot}}}\}$. Indeed, by comparing the signs of the first and last basis states on the RHS of~\eqref{eq:WembraceDjf2} with the same bases in middle matrix on the left we read off the very same update rules derived for two-dimensional space of Eqs.~\eqref{eq:WembraceDonehalf1} and~\eqref{eq:WembraceDonehalf2} for $n_{\text{tot}}=1$ (spanned by $\{\ket{1,0},\ket{0,1}\}$). By this step we reached the goal of completely characterizing the quantum-mechanical behavior of the $N\to N$ GMZI: Choosing any $n_i,n_j\geq0$ such that $n_i+n_j=n_{\text{tot}}$ we  generalize~\eqref{eqs:switchAction}
\begin{align}\label{eq:switchActGeneral}
  \euS_{\bphi}(\ket{0,\dots,n_i,\dots,n_j,\dots,0})
   &= (-)^{n_i\bphi(N)/\pi}\ket{0,\dots,n_{\s^{-1}(i)},\dots,0}\ox(-)^{n_j\bphi(N)/\pi}\ket{0,\dots,n_{\s^{-1}(j)},\dots,0} \nn\\
   &= (-)^{n_{\text{tot}}\bphi(N)/\pi}\ket{0,\dots,n_{\s^{-1}(i)},\dots,n_{\s^{-1}(j)},\dots,0}
\end{align}
and exactly the same expression as in Eq.~\eqref{eq:inputStateTrans} follows by induction.

Before our final example we mention that by the linearity of GMZI and the fact that the multimode Fock states in Eq.~\eqref{eq:multiPhotonBasis} form a complete basis of the $d$-dimensional Fock space of $N$ bosons it follows that the GMZI acts as switch for any input boson state -- not necessarily a Fock basis state. This generalizes the well-known behavior of an ordinary MZI, where if  an input state is, for instance, a single-mode coherent state $\ket{\a}$ it will evolve into $\ket{-\a}$ for a suitably chosen phase.

\begin{exa}
Let's show an example of the permutation sign derivation for $n_{\text{tot}}=2,3$ and $N=4$. It's instructive to perform the calculation over the whole Hilbert space for $n_{\text{tot}}=2$ and  only using the pivot subspace for $n_{\text{tot}}=3$. Eq.~\eqref{eq:SU2TotCarSpace} dictates the following decomposition for the first case: $\Hcal\simeq\Hcal_{1/2}\ox\Hcal_{1/2}\oplus\wp[{\Hcal_{1}\ox\Hcal_{0}}]$, where $\dim{\Hcal}=10$ is in agreement with the decomposition $4+3\times2$. The presence of the reducible representation ${1\over2}\ox{1\over2}$ requires extra attention since we don't know how it acts on $D$. It is always easy to find it on a one-to-one basis, as we will see, but the generic behavior for any $n_{\text{tot}}$ and $N$ is not immediately obvious. The spanning basis of $\Hcal$ is listed in Fig.~\ref{tab:PermSign} and the reducible space ${1\over2}\ox{1\over2}$  for $\ell=1$ is spanned by $\{\ket{0011},\ket{0110},\ket{1001},\ket{1100}\}$. For the phase choice $\bphi=(0\pi\pi0)$ we get the first line of the table and given the spanning basis we find
$$
\big(W^\dagger_{1,3}(1/2,\pi/2)\ox W^\dagger_{2,4}(1/2,\pi/2)\big)\diag{[-1,1,1,-1]}\big(W_{1,3}(1/2,\pi/2)\ox W_{2,4}(1/2,\pi/2)\big)=-\di_{4},
$$
which are the four blue $-1$'s in the $\ell=1$ row. The red 1's comes from the pivot space following~\eqref{eq:WdgDW0pipi}. The spanning basis ${1\over2}\ox{1\over2}$  for $\ell=2$ is $\{\ket{0101},\ket{0110},\ket{1001},\ket{1010}\}$ and so we calculate
$$
\big(W^\dagger_{1,2}(1/2,\pi/2)\ox W^\dagger_{3,4}(1/2,\pi/2)\big)\diag{[1,-1,-1,1]}\big(W_{1,2}(1/2,\pi/2)\ox W_{3,4}(1/2,\pi/2)\big)=\di_{4}
$$
to become the blue 1's in the $\ell=2$ row. The pivot space calculation follows (the red 1's) and this is how we obtained the positive permutation sign. Note that consistency requires all the signs of the last row to be identical.
\begin{center}
    \renewcommand{\arraystretch}{1.2}
    \extrarowheight=\aboverulesep
    %\addtolength{\extrarowheight}{\belowrulesep}
    \aboverulesep=0pt
    \belowrulesep=0pt
    \begin{figure}[h]
           \begin{tabular}{@{}>{\columncolor{white}[0pt][\tabcolsep]}  *{11}c @{}}
           \toprule
            {\ basis }  & $0011$ &  $0101$  &  $0110$ &  $1001$  & 1010 & 1100 & 2000 & 0200 & 0020 & 0002\\
                        \midrule
             {\cellcolor{lightgray}\ $(0\pi\pi0)$}  &  $-1$  &  $-1$ &  1   &1  & $-1$ & $-1$ & 1 & 1 & 1 & 1  \\
             {\cellcolor{lightgray}\ $\ell=1,j=\color{blue}{{1\over2}^{\ox2}},\color{red}{1}$}  &  \color{blue}{$-1$}  & \color{red}{1} &
            \color{blue}{$-1$} &
            \color{blue}{$-1$}  & \color{red}{1} & \color{blue}{$-1$} &\color{red}{1} &\color{red}{1} &\color{red}{1} &\color{red}{1}  \\
             {\cellcolor{lightgray}\ $\ell=2,j=\color{blue}{{1\over2}^{\ox2}},\color{red}{1}$}  & \color{red}{1} &  \color{blue}{1} &

            \color{blue}{1}
            &  \color{blue}{1} & \color{blue}{1} & \color{red}{1} &\color{red}{1} &\color{red}{1} &\color{red}{1} &\color{red}{1}  \\
             \bottomrule
             \hline
            \end{tabular}\\ \vskip .3cm
    \caption{An illustration of the example below~Eq.~\eqref{eq:switchActGeneral} showing how the overall phase emerges. Given the total photon number $n_\mathrm{tot}=2$ in $N=4$ modes the columns are the spanning basis states of the 10-dimensional boson Hilbert space $\Hcal$. The first row is the action of the chosen phase configuration $\bphi=(0\pi\pi0)$, the second row is the action of the first innermost BS layer $\ell=1$ (conjugation by the $W$ matrices) and the third row shows the action of the second layer $\ell=2$ of BSs (final in this case). The blue and red color indicates the action on the spin-$j$ subspaces of $\Hcal$ (their spanning bases differ for different layers). The final phase must be unique as we witness in the last row.}
    \label{tab:PermSign}
    \end{figure}
\end{center}
For $n_{\text{tot}}=3$ we get from~\eqref{eq:SU2TotCarSpace} $\Hcal\simeq\wp[\Hcal_{3/2}\ox\Hcal_0]\oplus\wp[\Hcal_{1}\ox\Hcal_{1/2}]$, where $\dim{\Hcal}=20=4\times2+6\times2$. Choosing only a portion of the pivot space $\Hcal_{3/2}$ we obtain for, for example, $\bphi=(0\pi0\pi)$ the negative permutation sign by following the updates rules in Eqs.~\eqref{eqs:simplifiedConjugation} as documented in Fig.~\ref{tab:PermSignPivot}.
\begin{center}
    \renewcommand{\arraystretch}{1.2}
    \extrarowheight=\aboverulesep
    %\addtolength{\extrarowheight}{\belowrulesep}
    \aboverulesep=0pt
    \belowrulesep=0pt
    \begin{figure}[h]
           \begin{tabular}{@{}>{\columncolor{white}[0pt][\tabcolsep]}  *{6}c @{}}
           \toprule
            {\ basis } & $\cdots$ & 3000 & 0300 & 0030 & 0003\\
                        \midrule
             {\cellcolor{lightgray}\ $(0\pi0\pi)$}  &   $\cdots$  &  $1$ &  $-1$   & $1$  & $-1$  \\
             {\cellcolor{lightgray}\ $\ell=1$}      &   $\cdots$  &   $1$ &  $-1$   & $1$  & $-1$   \\
             {\cellcolor{lightgray}\ $\ell=2$}      &   $\cdots$  &  $-1$ &  $-1$   & $-1$  & $-1$  \\
             \bottomrule
             \hline
            \end{tabular}\\ \vskip .3cm
    \caption{The same example as in Fig.~\ref{tab:PermSign} for a different phase configuration and for $n_\mathrm{tot}=3$, where it would start to be tedious to write down the whole basis. The overall phase can be, however, obtained just by `evolving' the highest-weigh basis labeling the columns.}
    \label{tab:PermSignPivot}
    \end{figure}
\end{center}
\end{exa}

\section{Quantitative benchmarking of resource scaling}\label{sec:benchScale}

In order to show the asymptotic scaling of the relevant resources of our GMZI-based architecture we will construct a toy model of quantum architecture based on the rotated surface code and study how the number of switches and their (optical) depth scales under reasonable assumptions. For concreteness, our quantum computer will be tasked with implementing  a transversal CNOT between remote logical patches with the help of switches. This is not a universal quantum architecture but it forms a crucial part of logical information exchange using advanced logical operations (such as lattice or code surgery~\cite{cowtan2024css,horsman2012surface}) and even incorporation of magic state factories for more complete (universal) architectures. So it is a toy, but not trivial, model of distributed quantum architecture.

Our modular fault-tolerant architecture consists of~$n$ surface-code patches, each containing $N$ physical qubits, which are interconnected by optical routers so that any pair of patches can execute a logical CNOT transversally and in parallel, i.e., by $N$ simultaneous pairwise physical CNOTs (or photon-mediated entangling operations) between corresponding pairs of logical qubits. The hardware primitive is a switch (router) with $N$ inputs, $kN$ outputs ($k$ and $N$ fixed so that we cannot just make the switch bigger and keep the parameters intact which is a serious manufacturing constraint we follow), and constant optical depth $\d$ ($\d=1$ for our GMZI-based construction). One such router connects a patch to $k$ others. When $n$ grows and the reachable fraction of patches $k_{\mathrm{eff}}/n$ is held constant by concatenating routers into $k$-ary fan-out trees we show  the growth of the number of routers per patch  and the end-to-end optical depth scaling. For a rotated surface code patch of distance~$d$, the transversal CNOT couples the $N=d^2$ data qubits of the control patch pairwise to the corresponding data qubits of the target patch. Because the surface code is CSS, bitwise physical CNOT implements logical CNOT exactly. In an optically interconnected implementation, executing this gate between different patches requires the simultaneous delivery of $N$ optical channels from (or between) the two patches -- either~$N$ photons that directly mediate the gate or photons that herald~$N$ Bell pairs consumed by gate teleportation~\cite{barrett2005efficient,browne2005resource}.

The scaling question is as follows: Assume $k$ fixed. Then as $n$ increases,  single router's  connectivity ratio $k/n$ decays. The architecture under study restores it by \emph{concatenation} as mentioned on page~\pageref{para:scaling}: each of the $k$ output blocks of a router  feeds the $N$-port input of a child router, forming a $k$-ary tree. A tree with $t$ levels terminates in $k^t$ output blocks, i.e., reaches $k^t$ candidate destination patches. The design requirement is
\begin{equation}\label{eq:requirement}
\rho\leq\frac{k_{\mathrm{eff}}(n)}{n}
\end{equation}
for all $n$, where $\rho\df{k/n_0}$ ($n_0$ patches with one GMZI-based switch and assuming $k<n_0$) and $k_{\mathrm{eff}}(n)=k^{t(n)}$ is the per-patch reach. How do the number of routers and the overall optical depth scale with $n$?

Let $t(n)=\ceil{\log_k{[\rho n]}}$. The fan-out tree of one patch contains the following number of switches:
\begin{equation} \label{eq:NR}
  N_R(n)=\sum_{\ell=0}^{t(n)-1} k^{\ell}={k^{t(n)}-1\over k-1}.
\end{equation}
Then
\begin{equation}%\label{}
  {\rho n - 1\over k-1}\leq N_R(n)\leq{k\rho n - 1\over k-1}
\end{equation}
and hence $N_R(n) = \Theta(n)$, i.e., linear in~$n$ with a slope between $\rho/(k-1)$ and $k\rho/(k-1)$. Summed over all $n$ patches we get
\begin{equation}\label{eq:Ntot}
N_\mathrm{tot}(n)=n\,N_R(n)\approx{\rho\over k-1}\,n^2=\Theta(n^2).
\end{equation}
The most important parameter is, however, the total optical depth. A photon traversing one fan-out tree passes $t(n)$ switches in series so the one-way optical depth is
\begin{equation}\label{eq:depth}
D(n)=\d t(n)=\d\ceil{\log_k{[\rho n]}}=\Theta(\log n).
\end{equation}
With trees at every receiving patch, the total number of switches is $2N_\mathrm{tot}(n)\approx{2\rho\over k-1}n^2$ routers and end-to-end depth $2D(n)=2\d\ceil{\log_k{[\rho n]}}$. The constant factors do not change the asymptotic: $\Theta(n^2)$ for hardware and $\Theta(\log n)$ for depth.

\begin{figure}[h]
  \resizebox{14.5cm}{!}{\includegraphics{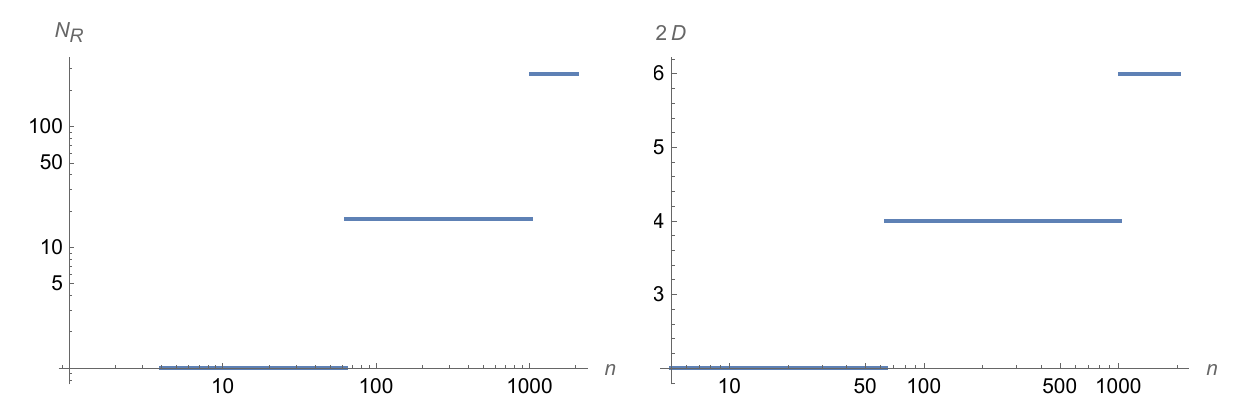}}
  \caption{Left: The number of GMZI switches per logical patch as a linear function of the logical qubit (patch) number~$n$ derived in~\eqref{eq:NR}. Right: The doubled optical depth $D(n)$ (Eq.~\eqref{eq:depth}) counting how many active optical elements a photon travelling from one patch to another encounters (in the direct scenario) or if two photons `meet' somewhere in the middle (the entanglement scenario). The staircase scaling is logarithmic and therefore showing a high scalability of our GMZI-based proposal.}
  \label{fig:scaling-example}
\end{figure}

We illustrate the scaling in Fig.~\ref{fig:scaling-example} for up to $n=1024$ logical qubits (1024 patches), $k=16,\rho=1/4$ and $\d=1$.
Nowhere in our analysis figures the actual physical qubit number~$N$ but as discussed in the main text an $N\to M\equiv Nk$ GMZI could get unrealistically big as the distance~$d=\sqrt{N}$ increases. This is easy to fix by using many smaller GMZIs in parallel per patch. In this way we preserve the depth $D(n)$ (the photons pass through exactly the same number of active elements) but the number of (inevitably smaller) $N'\to N'k$ GMZIs per patch linearly increases with~$N$.

As a concluding remark it is worth mentioning that our effort to keep $\rho$ constant is the worst case scenario in a certain sense. We could sacrifice the constant connectivity ratio for a slow-decreasing one with an increasing~$n$ and it will still not jeopardize the functionality and scalability of a fault-tolerant quantum computer. The compilation layer will make sure that the directly connected patches are prioritized for logical operations. This may not be always possible for an arbitrary quantum algorithm but then the penalty in terms of temporal and physical footprint is likely to be minimal. How minimal is certainly an interesting direction to pursue in the quantum architecture design.

%\bibliographystyle{unsrt}

%\bibliography{RefsIcon}

\end{document}